\documentclass{aa}  

\usepackage{txfonts}
\usepackage{wasysym,graphicx,verbatim,lscape,rotating,hyperref}

\newcommand{\bes}{Besan\c{c}on}
\newcommand{\euclid}{\emph{Euclid}}
\newcommand{\ts}{\textstyle}

\begin{document}

   \title{The microlensing rate and distribution of free-floating planets towards the Galactic bulge}


   \author{M. Ban \inst{1} \fnmsep \and
          E. Kerins \inst{1} \fnmsep \and
          A.C. Robin \inst{2} \fnmsep
          }

   \institute{Jodrell Bank Centre for Astrophysics, University of Manchester
              \label{inst1}
              \and
	        Institut Utinam, CNRS UMR6213, Universit\'{e} de Franche-Comt\'{e}, Observatoire de \bes{}
	        \label{inst2}
             }

   \date{Received Month date, year; accepted Month date, year}

 
  \abstract
   {Ground-based optical microlensing surveys have provided tantalising, if inconclusive, evidence for a significant population of free-floating planets (FFPs). Both ground- and space-based facilities are being used and developed which will be able to probe the distrubution of FFPs with much better sensitivity. It is also vital to develop a high-precision microlensing simulation framework to evaluate the completeness of such surveys.}
   {We present the first signal-to-noise limited calculations of the FFP microlensing rate using the \bes{} Galactic model. The microlensing distribution towards the Galactic centre is simulated for wide-area ground-based optical surveys ($I$ band) such as OGLE or MOA, a wide-area ground-based near-infrared survey ($K$ band), and a targeted space-based near-infrared survey ($H$ band) which could be undertaken with \euclid{} or \emph{WFIRST}.}
   {We present a calculation framework for the computation of the optical and near-infrared microlensing rate and optical depth for simulated stellar catalogues which are signal-to-noise limited, and take account of extinction, unresolved stellar background light, and finite source size effects, which can be significant for FFPs.}
   {We find that the global ground-based $I$-band yield over a central 200~deg$^2$ region covering the Galactic centre ranges from 20 Earth-mass FFPs year$^{-1}$ up to 3,500 year$^{-1}$ for Jupiter FFPs in the limit of 100\% detection efficiency, and almost an order of magnitude larger for a $K$-band survey. For ground-based surveys we find that the inclusion of finite source and the unresolved background reveals a mass-dependent variation in the spatial distribution of FFPs. For a targeted space-based $H$-band covering 2 deg$^2$, the yield depends on the target field but maximises close to the Galactic centre with around 76 Earth to 1,700 Jupiter FFPs per year. For near-IR space-based surveys like \euclid{} or \emph{WFIRST} the spatial distribution of FFPs is found to be largely insensitive to the FFP mass scale.}
   {}


   \titlerunning{Microlensing rate of free-floating planets}
   \authorrunning{M. Ban, E. Kerins \& A.C. Robin}
   \maketitle
%

\section{Introduction}		

With over 2000 confirmed exoplanets\footnote{\url{http://exoplanets.eu}} our knowledge of the statistics of the exoplanet population has grown enormously over the two decades since exoplanets were first identified. Radial velocity and transit techniques have been responsible for most of the detected exoplanets. Both of these techniques favour the detection of massive planets with small host separation and so our knowledge of lower mass planets further out from the host, especially beyond the ice line, remains highly incomplete. Currently, only eight exoplanets have been published with mass below $30 M_{\oplus}$ and host separation beyond 2 AU; seven of these have been detected using microlensing.

Microlensing currently offers the best promise for characterising the exoplanet population beyond the ice line down to Earth masses \cite[for a review see][]{gau12}. Favoured models of planet formation, such as the core accretion model, suggest that low-mass planets beyond the ice line typically do not migrate far from the orbits where they form \cite[e.g.][]{mor09}. The ability to compile statistics on cool low-mass planets beyond the ice line therefore provides a very important diagnostic of in situ planet formation. 

Microlensing is also able to detect planetary-sized bodies which are not bound to a host star. We refer to these objects by the commonly-used term of free-floating planets (FFPs), though such objects may not originally have formed as a conventional planet around a host star. \cite{sumi11} found an excess of short-duration microlensing events in the MOA-2 survey which was above what would be expected from lensing by the ordinary stellar and brown dwarf population. Their best-fit models to this excess requires a substantial population of Jupiter-mass FFPs with an inferred abundance of 1.8 per Galactic star. Isolated planetary-mass objects are also seen in near-infrared proper motion surveys of young stellar clusters, with the lowest mass examples being around 4--8 $M_J$ (e.g. \citealt{zap14} and references therein; for a review see \citealt{luh12}). \cite{ma16} computed the microlensing properties of a population of FFPs generated from core accretion simulations. They found that the predicted FFP timescales were generally too short to explain the excess of short events found by MOA (\citealt{sumi11}). This could suggest that the FFP population detected by MOA formed as isolated planet-sized objects rather than as part of initially bound planetary systems.

Microlensing computations are often undertaken using analytic or semi-analytic models for the Galactic density and velocity distribution, together with scaling laws for the source luminosity and mass functions. These prescriptions range in their sophistication but this approach generally does not enforce consistency across the multiple interrelated pieces of input physics. Furthermore, these models usually ignore extinction or employ a simple prescription such as an extinction ``screen'' or a uniform extinction model. 

Whilst this approach has proved sufficient for comparison with modest samples of the order of 100 events, surveys are now providing completeness-corrected samples of several hundred to 1000 events \citep{sumi13,wyrzykowski15}. As the catalogue of detected microlensing events over the last 20 years now surpasses $10^4$ events, with a current detection rate of around 2000 events per year, our understanding of this data is now firmly limited by the sophistication of our theoretical models.

The use of synthetic population synthesis models \citep{robin03, robin12} for microlensing calculations provides a promising approach to interpreting large-scale microlensing datasets \citep{ker09,awip16}. By the same token this approach can be used to undertake detailed optimisation studies for future microlensing survey programmes \citep{penny13}. Microlensing data is already challenging these models and therefore has an important role to play in improving our understanding of the structure of the inner Galaxy. 

In the current paper we look at the expected microlensing signature of a population of FFPs using the \bes{} Galaxy population synthesis model \citep{robin03, robin12}. Whilst FFP microlenses are observable as short-duration single-lens events, they are expected to be commonly affected by finite source size effects and therefore their spatial distribution will, to some extent, be affected by differences in stellar population along different sight lines. Additionally, whilst the microlensing rate is optimised towards the Galactic Centre the effects of dust and of the background Galactic surface brightness due to unresolved stars becomes more severe, with the severity dependent on observation wavelength. Population synthesis modelling provides a suitable framework for assessing all of these effects. 

The potential for spatial systematics in the FFP microlensing distribution is particularly relevant for possible space-based exoplanet microlensing surveys which may occur within the next decade. The Exoplanet Euclid Legacy Survey (ExELS) is a proposal for an additional science survey on the ESA Euclid mission \citep{penny13,mcd14}. ExELS would monitor a region of 1.6 deg$^2$ close to the Galactic Centre in the $H$ band. Euclid has been selected by ESA for launch around 2019 \citep{lau11}, though additional science programmes are yet to be confirmed. A larger microlensing programme forms part of the core science activity for the proposed NASA WFIRST mission \citep{sper13}. Both surveys would be conducted over relatively small fields of view and so it is important to understand how the choice of survey location may affect FFP sensitivity. The purpose of this paper is to assess this and to compare our data with yields expected from current generation ground-based optical microlensing surveys such as OGLE, MOA, and the recently initiated KMTNet survey, which survey wide regions of the inner Galaxy. We also consider a ground-based wide-area near-infrared survey which could plausibly be undertaken within the next few years with a facility such as VISTA.

The paper is laid out as follows. In Section \ref{sec:bes} we outline the version of the \bes{} model used for our microlensing computations. In Section \ref{sec:microsim} we describe in detail how we carry out microlensing simulations with the \bes{} model. Our simulation results for Earth- to Jupiter-mass FFPs are presented in Section \ref{sec:res} for wide-area optical and near-IR surveys on the ground as well as a targeted near-IR space-based survey. Signal-to-noise limited maps of optical depth, rate, and event duration are presented which take account of finite source size effects and the effect of unresolved stars on sensitivity. We conclude with Section \ref{sec:conc}.

\section{The Besan\c{c}on model of the Galaxy}	\label{sec:bes}

We use version 1112 of the \bes{} model which comprises four principal stellar populations: a thin and thick disk, a spheroid, and a bulge. A similar model (1106) was used by \cite{penny13}, and the main update for version 1112 is the bulge structure as tested by \cite{robin12}. We refer the reader to those papers for additional model details. We list here the most relevant characteristics. 

The thin disk density distribution is a density law following Einasto's ellipsoids (expression given in \citealt{robin03}) with  a central hole. The family comprises seven subcomponents of various ages ranging from 0--10 Gyr. Each population has self-consistent kinematics with the older populations having increasing scale height and, therefore, increasing velocity dispersion. The Local Standard of Rest (LSR) speed is 226 km s$^{-1}$. The thin disk stars are generated with a mass function $dn/dm \propto m^{-x}$, with $x = 3$ above $1 M_{\odot}$ and $x = 1.6$ below this down to the hydrogen burning limit, as determined from the Hipparcos luminosity function.

The bulge density law is taken to be a triaxial Gaussian bar with scale lengths of 1.46:0.49:0.30 kpc, a position angle of $12.^{\circ}89$ and a cut-off radius of 3.43 kpc. The model also includes a second much smaller bar component, orientated with a similar position angle but with scale lengths of 4.44:1.31,0.80 kpc. However, this component has very low mass and therefore negligible impact on our microlensing results. Both bar components have a velocity dispersion $(U,V,W) = (131,106,85)$ km s$^{-1}$. Overall the model is described as the ``S+E'' model in Table 2 of \cite{robin12}. The bulge IMF is a power law with slope $x = -2.35$ above 0.7 M$_{\odot}$.

The thick disk and spheroid populations, whilst contributing minimally to the microlensing rates computed in this paper, are included and are as described in \cite{robin12}. In the model the Sun is located 15 pc above the mid-plane at a distance of 8 kpc from the Galactic Centre. The solar apex motion is $(U , V ,W) = (10.3, 6.3, 5.9)$ km s$^{-1}$ with respect to the LSR.

\section{Microlensing event simulation} \label{sec:microsim}

We wish to simulate microlensing maps in three different contexts: i) a simulation of current wide-field ground-based optical surveys ($I$-band) such as OGLE \citep{wyrzykowski15} or MOA \citep{sumi13}; ii) a simulation of a ground-based wide-field near-infrared survey ($K$-band) like the VISTA VVV survey (\citealt{saito12}), but with a higher cadence and deeper sensitivity suitable for microlensing FFP detection; and iii) a narrow-field space-based near-infrared survey ($H$-band), such as ExELS, which has recently been proposed as an additional exoplanet science survey for \euclid{} \citep{penny13,mcd14}.

Our simulations draw on artificial stellar catalogues generated with the \bes{} population synthesis model described in Section \ref{sec:bes}. We use an adaptive simulation solid angle to ensure good statistics across the whole magnitude range whilst keeping computational time to a minimum.
To this end we subdivide the simulations into four catalogues (A--D) spanning different magnitude ranges (Table \ref{tab:catals}). For more numerous fainter sources we restrict the simulations to narrower solid angles in order to provide roughly comparable statistics in each catalogue. The ratios of the solid angles are used to normalise all results to a common field of view. The chosen parameters provide around 2500 sources for each catalogue towards Baade's window ($l = 1\degr ,  b = -4\degr$), with declining or increasing numbers moving away or towards the Galactic Centre, respectively. We perform computations for 3321 cells spanning an area of 200~deg$^2$, giving a spatial resolution of $15\times15$~arcmin$^2$ per cell. 
\begin{table}
 \centering
 \begin{minipage}{140mm}
  \caption{Besan\c{c}on catalogue parameters adopted for this work.}
\label{tab:catals}
  \begin{tabular}{@{}lp{1.5in}p{1.0in}@{}}
  \hline
  Main band & $K$-band \\ 
  colour bands & $I-K$, $J-K$, $H-K$ \\ \hline
  target area [deg] & $-10\leq l \leq 10$, $-5\leq b \leq5$\\		
  area per cell [deg$^2$] & $0.25\times0.25$\\
  distance range [kpc] & 0-15\\ \hline
  magnitude range & ctlg.A : K = 0-12 \\ 
   & ctlg.B : \hspace*{5.5mm}12-16 \\
   & ctlg.C : \hspace*{5.5mm}16-20 \\
   & ctlg.D : \hspace*{5.5mm}20-99 \\ \hline
  solid angle [deg$^2$] & ctlg.A : 0.0625 \\
   & ctlg.B : $6.8\times 10^{-3}$ \\
   & ctlg.C : $2.1\times 10^{-4}$ \\
   & ctlg.D : $3.6\times 10^{-5}$ \\ \hline
\end{tabular}
\end{minipage}
\end{table}

For each cell the simulation proceeds by treating every catalogue star as both a potential lens or source. As a source the microlensing contribution of all catalogue stars closer to the observer is computed, weighting appropriately for their lensing cross section, motion relative to the source, source brightness and catalogue solid angle (see Section \ref{subsec:microvals}). The final results are averaged over all potentially detectable events according to signal-to-noise criteria which we now discuss.

\subsection{Microlensing detectability}	\label{subsec:detect}

The microlensing rate for a given survey is fundamentally limited by signal-to-noise considerations. The signal here refers to the source star brightness which must at least be detectable to a survey whilst at maximum microlensing magnification. The noise is assumed to be dominated by photon noise from the event itself, the sky background, and the unresolved stellar background flux. We therefore write the signal-to-noise at time $t$ as  
\begin{equation}
S/N(t) = \frac{10^{0.2 m_{\rm zp}} \, t^{1/2}_{\rm exp} \, A(t) \, 10^{-0.4m_*}}{\sqrt{10^{-0.4 m_{\rm stars}} + \Omega_{\rm psf}10^{-0.4\mu_{\rm sky}} + A(t) \, 10^{-0.4m_*}}},
\label{eq:sn}
\end{equation}
where $m_*$ is the apparent magnitude of the source star, $A(t)$ is the microlensing magnification, $\mu_{\rm sky}$ is the sky surface brightness, and $\Omega_{\rm psf} = \pi \theta_{\rm psf}^2/4$ is the solid angle of the survey point spread function (PSF), with $\theta_{\rm psf}$ the PSF full width at half-maximum size. The combined magnitude contribution of all unresolved sources at the location of the microlensing source is $m_{\rm stars}$, whilst $m_{\rm zp}$ and $t_{\rm exp}$ are the survey-dependent zero-point magnitude and exposure time, respectively. The values we adopt for our three survey configurations are provided in Table \ref{tab:survs}. The survey-dependent parameter values we choose for $m_{\rm zp}$ and $t_{\rm exp}$  reflect the current capability of $I$-band surveys such as OGLE and MOA, the capability of a deep $K$-band survey conducted with a telescope like VISTA, and a space-based $H$-band survey, such as the proposed \euclid{} ExELS survey. For the ground-based surveys our choices for $m_{\rm zp}$ and $t_{\rm exp}$ ensure $4\%$ photometric precision at a limiting magnitude $m_{\rm lim} = m_{\rm zp}$, assuming the flux is dominated by the source.  The limiting sensitivity of our $K$-band ground-based survey is set to be similar to that of the $H$-band space-based survey in order to provide a direct comparison of the advantages of observing from space. The true values of $m_{\rm zp}$ and $t_{\rm exp}$ for a particular survey are unimportant for our calculations provided they imply a similar photometric precision at $m_{\rm lim}$.

Two elements of equation (Eq.\ref{eq:sn}) require further clarification. Firstly, we must specify precisely what we mean by unresolved stars, which contribute to $m_{\rm stars}$. For this we can employ another signal-to-noise condition which compares the signal due to the PSF flux of a star $j$, with magnitude $m_j$, to the combined noise contribution due to flux within the PSF from all simulated stars fainter than $m_j$: 
\begin{equation}
S/N_j = \frac{10^{0.2 m_{\rm zp}} \, t^{1/2}_{\rm exp} \, 10^{-0.4m_j}}{\sqrt{\frac{\ts \Omega_{\rm psf}}{\ts \Omega_{\rm cat}} \sum\limits_{m_i > m_j} 10^{-0.4m_i}}},	
\label{eq:resstar}
\end{equation}
where $\Omega_{\rm cat}$ is the solid angle of the simulation catalogue. Since we employ multiple catalogues of varying solid angle (given in Table \ref{tab:catals}) there is also an implied summation over star catalogues in Eq.\ref{eq:resstar}. Sorting the catalogues by stellar magnitude we find the faintest star $j$ which still provides $S/N_j > 3$ with respect to the PSF noise contribution from the remaining fainter sources. These fainter sources constitute the unresolved stellar background at the location of source $j$ and their combined flux, expressed as a magnitude, defines $m_{\rm stars}$.

The second element of Eq.\ref{eq:sn} which we must specify is the microlensing magnification factor $A$ required for a detection. We define a detection to occur when $S/N > 50$ at peak magnification and use Eq.\ref{eq:sn} to solve for the minimum magnification $A_{\rm min}$. 
\begin{table*}
 \centering
 \begin{minipage}{140mm}
  \caption{Assumed survey parameters for ground-based $I$- and $K$-band wide-area surveys, as well as an $H$-band space-based survey. For the ground-based $I$ and $K$-band surveys $t_{\rm exp}$ is set so that $m_{\rm zp}$ corresponds to the survey limiting magnitude for $4\%$ photometric precision. The values for the $H$-band space-based survey are based on the propsed \euclid{} ExELS survey and are taken from Table 2 of \cite{penny13}.}
\label{tab:survs}
  \begin{tabular}{@{}lp{1.0in}p{1.0in}p{1.0in}@{}}
  \hline
  & $I$ band (Ground) & $K$ band (Ground)& $H$ band (Space)\\ \hline
  $u_{\rm max}$ & 1 & 1 & 1 and 3\\
  $\mu_{\rm sky}$ [mag/arcsec$^2$] & 19.7 & 13.0 & 21.5 \\	
  $\theta_{\rm psf}$ [arcsec] & 1.0 & 1.6 & 0.4 \\
  $m_{\rm zp}$ & 20.4 & 24.6 & 24.9 \\
  $t_{\rm exp}$ [secs] & 180 & 60 & 54 \\
\end{tabular}
\end{minipage}
\end{table*}
The factor $A_{\rm min}$ translates into a constraint on the maximum (threshold) impact parameter required for detectability, $u_{\rm t}$. When computing microlensing statistics we must also specify a largest impact parameter $u_{\rm max}$ which defines the boundary of what we accept as an event, so that events are counted only where the impact parameter falls within the range $0 < u < u_{\rm max}$. The provided $u_{\rm t} > u_{\rm max}$ events are detectable for all allowed impact parameters. However if $u_{\rm t} < u_{\rm max}$ then only a fraction of allowable events are detectable and so their contribution to the event rate must be weighted by $u_{\rm t}/u_{\rm max}$, whilst their contribution to the optical depth is weighted by the event cross section $(u_{\rm t}/u_{\rm max})^2$. For ground-based surveys it is standard to adopt $u_{\rm max} = 1$, corresponding to a minimum magnification of 1.34. For a space-based survey, where atmospheric effects are not an issue and greater photometric stability is possible, we may expect to be able to reliably detect fainter variations. In our simulation our default condition for space-based detection is also $u_{\max} = 1$ since it allows us to compare our data with ground-based numbers under like-for-like assumptions. In addition, we consider a maximum impact parameter $u_{\rm max} = 3$ for space-based detection, corresponding to a minimum magnification factor of 1.02. 

Since the \bes{} catalogue supplies the radius and distance of all simulated stars we incorporate finite source corrections where necessary in computing $u_{\rm t}$, assuming a uniform source brightness profile.

\subsection{Discrete microlensing computations} \label{subsec:microvals}

\cite{ker09} provided the first microlensing calculations from a Galactic population synthesis simulation. Whilst they described the calculation steps the relevant formulae for computing the optical depth and average event timescale from discrete simulated stellar catalogues were not explicitly given and so we provide them here. We also generalise their calculations to include finite source size effects which form a significant correction for FFP microlensing. 

\subsubsection{Optical depth} 

Given a definition for the event detectability threshold limit $u_{\rm t}$ (see Section \ref{subsec:detect}), we can express the source-averaged microlensing optical depth for events with impact parameters $u < u_{\rm max}$ as:
  \begin{equation}
    \tau = \left( \sum_s \frac{\ts 1}{\ts \Omega_{{\rm cat},s}} \sum_j U_j^{(2)}\right)^{-1} u_{\rm max}^2 \sum_s \frac{\ts 1}{\ts \Omega_{{\rm cat},s}} \sum_j U_j^{(2)} \sum_l \sum_{i,D_i < D_j}  \frac{\ts \pi  \theta_{{\rm E},ij}^2}{\ts \Omega_{{\rm  cat},l}}.
    \label{eq:tau}
    \end{equation}
The summations in Eq.\ref{eq:tau} are performed over lensing stars $i$ drawn from star catalogue $l$ and source stars $j$ drawn from star catalogue $s$. The innermost summation is over all lenses at distances $D_i$ which are closer than the source distance $D_j$. Eq.\ref{eq:tau} allows for multiple source and lens catalogues with varying solid angle $\Omega_{\rm cat}$. The angular Einstein radius $\theta_{\rm E}$ is given by
  \begin{equation}
    \theta_{{\rm E},ij} = \sqrt{\frac{4G M_i (D_j - D_i)}{c^2 D_j D_i}},
    \label{eq:theta}
  \end{equation}
where $M_i$ is the lens mass and $G$ and $c$ take their usual meaning. The optical depth must be weighted over the fraction of sources which involve detectable events. To this end we define the impact parameter weighting factor
 \begin{equation}
   U_j^{(N)} = \frac{\ts \sum_l \Omega_{{\rm cat},l}^{-1} \sum_{i,D_i<D_j} \mbox{min}[1,(u_{{\rm t},ij}/u_{\rm max})^N]}{\ts \sum_l \Omega_{{\rm cat},l}^{-1} \sum_{i,D_i<D_j} 1},
   \label{eq:uwgt}
   \end{equation}
which is the lens-averaged expectation value of $(u_{\rm t}/u_{\rm max})^N$ for source $j$, where the ratio is capped at a maximum value of unity. For the optical depth we set $N = 2$ as $\tau$ scales as $u_{\rm max}^2$. The lens averaging of $U_j^{(N)}$ is unnecessary in the strict point source limit where $U_j^{(N)} \rightarrow \min[1,(u_{{\rm t},j}/u_{\rm max})^N]$, with $u_{{\rm t},j}$ depending only on source brightness. However, lens averaging is required when finite source size corrections become important, which is generally the case for FFP microlensing.

\subsubsection{Event rate and average duration} 

The event duration is defined by the Einstein radius crossing time: $t_{\rm E} = \theta_{\rm E}/\mu$, where $\mu$ is the lens--source relative proper motion. The \bes{} catalogues we generate include stellar proper motions, allowing us to easily compute $\mu$, and hence $t_{\rm E}$, for each lens--source pair. 

In computing the {\em average}\/ event duration we must ensure that it is rate-weighted ({\it i.e.} weighted in proportion to the microlensing event rate at each duration). Since the event rate contribution due to lens $i$ and source $j$ scales as
  \begin{equation}
  W_{ij} = u_{{\rm t},ij} D_i^2 \mu_{ij} \theta_{{\rm E},ij},
    \label{eq:wrate}	
  \end{equation}
where $\mu_{ij}$ is the relative lens--source proper motion, the rate-weighted average event duration is
{\scriptsize
  \begin{equation} 
    \langle t_{\rm E} \rangle = \left( \sum_s 
    \frac{\ts 1}{\ts \Omega_{{\rm cat},s}} \sum_j \sum_l 
    \sum_{i,D_i < D_j} \frac{W_{ij}}{\ts \Omega_{{\rm cat},l}} \right)^{-1} 
    \sum_s \frac{\ts 1}{\ts \Omega_{{\rm cat},s}} \sum_j \sum_l \sum_{i,D_i < D_j} \frac{\ts W_{ij}}{\ts \Omega_{{\rm  cat},l}} \frac{\ts \theta_{{\rm E},ij}}{\mu_{ij}}.
    \label{eq:time}
  \end{equation} 
}
The source-averaged event rate $\Gamma$ is then obtained in straightforward fashion: 
\begin{equation}
   \Gamma = \frac{2}{\pi} \frac{\tau}{\langle t_E \rangle}.
    \label{eq:eventrate}
\end{equation} 
   Finally, in order to compute the total event rate within a given patch of sky we need to compute the effective number of detectable source stars per unit sky area:
   \begin{equation}
     N_* = \sum_s \frac{\ts 1}{\ts \Omega_{{\rm cat},s}} \sum_{j,U_j^{(1)} > 0} 1,
   \label{eq:sourcenum}
   \end{equation}
where $U_j^{(1)}$ is given by Eq.\ref{eq:uwgt} with $N = 1$ and ensures that we sum only over sources which are bright enough to make a non-zero contribution to the rate. The total event rate per unit sky area containing $N_*$ effective sources is then $\Gamma\times N_*$.

\subsection{Simulations} \label{subsec:sims}

The simulations are performed over 3321 lines of sight covering a total area of 200 deg$^2$ spanning $|b| < 5\degr$ and $|l| < 10\degr$. \bes{} star catalogues are generated over a grid of positions with an angular resolution of $15\times15$ arcmins, corresponding to the limiting resolution of the 3D dust map. The catalogues provide a wide range of stellar parameters. The parameters we use for the microlensing calculations are stellar distance, $K$-band apparent magnitude along with $I-K$ and $H-K$ colours, stellar proper motions in $l$ and $b$, and stellar mass and radius (the latter is required for finite source size corrections to $u_{\rm t}$).

\begin{figure}[bp]		
\hspace*{-0.2in}
\includegraphics[width=3.8in]{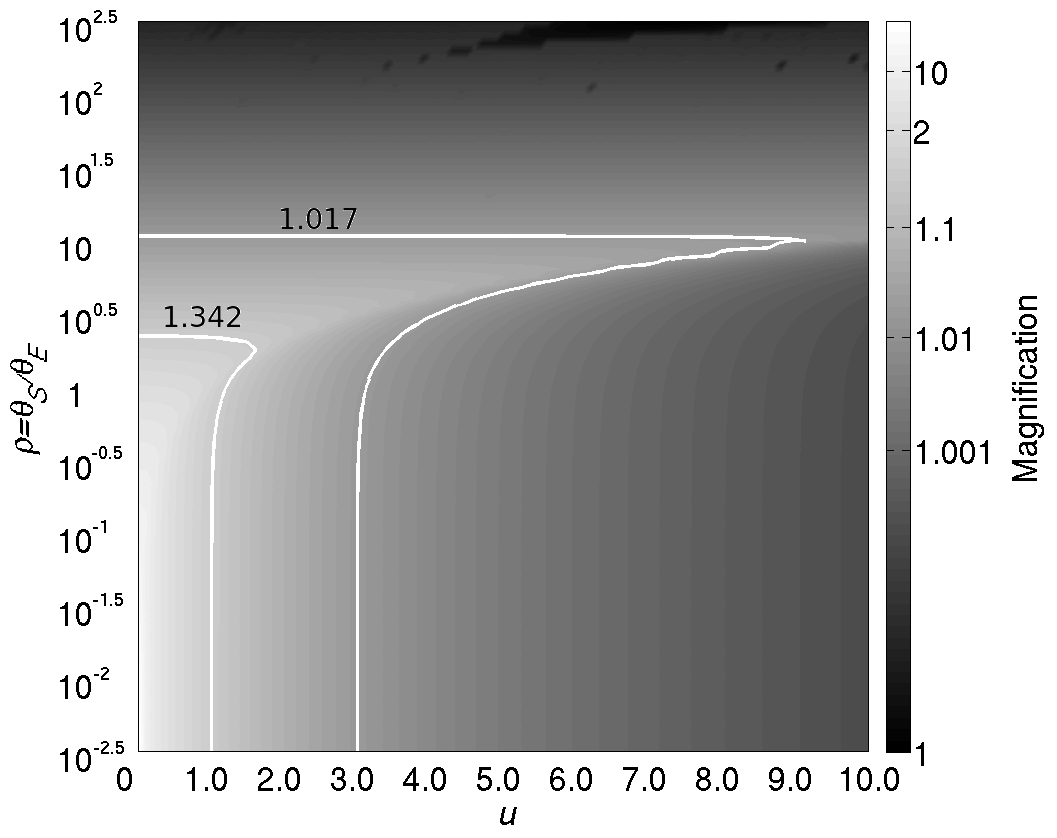}
\caption{Finite source magnification with $u$ on the x-axis and $log_{10}\rho$ on the y-axis, where $\rho$ is the angular radius of a source in units of $\theta_{\rm E}$. The two white curves correspond to our two adopted thresholds, $u_{\rm max}=1$ and $u_{\rm max}=3$.}
\label{fig:amp}
\end{figure}

\begin{figure}
{\centering	
\hspace*{-0.3in}
(a) Ground-based $I$-band\\
\includegraphics[width=3.8in]{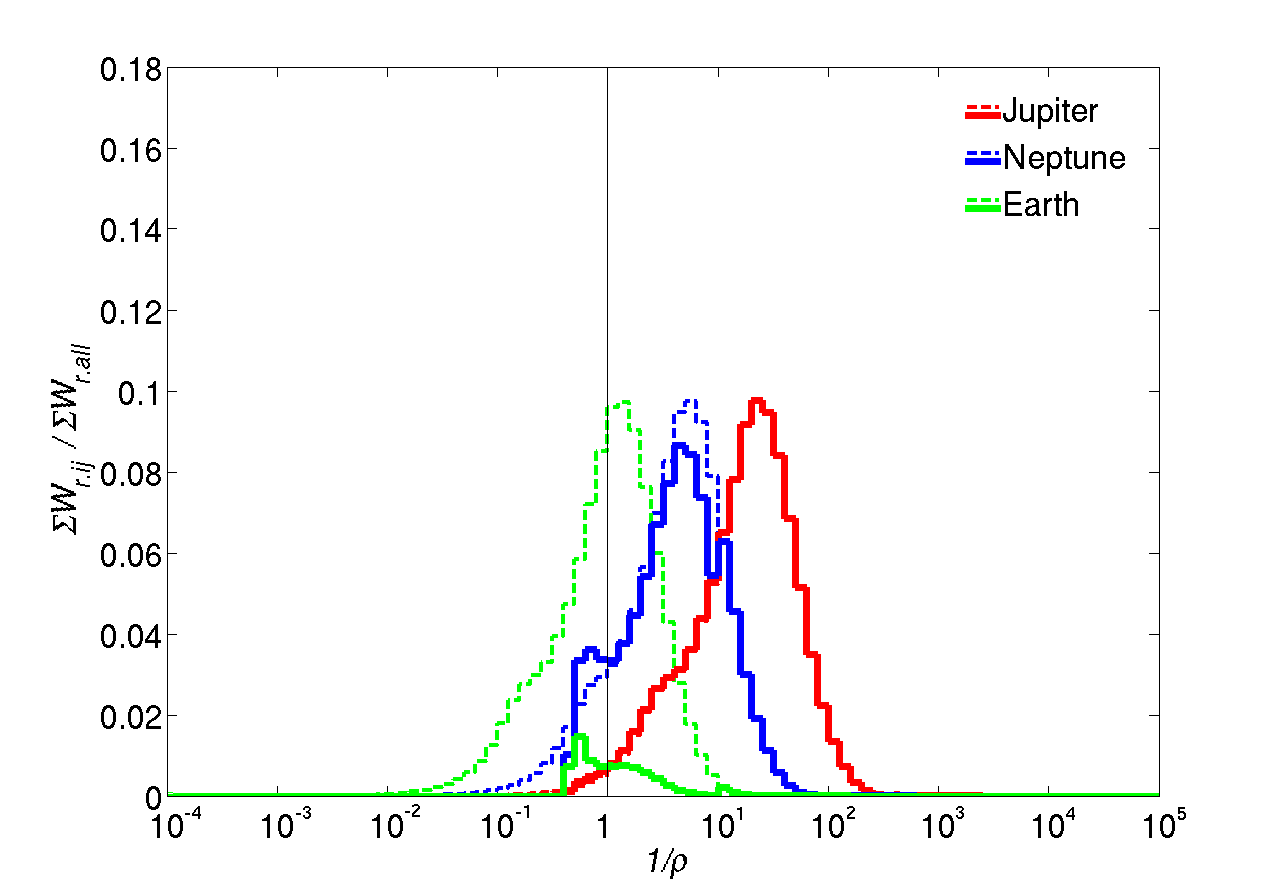}\\
\hspace*{-0.3in}
(b) Ground-based $K$-band\\
\includegraphics[width=3.8in]{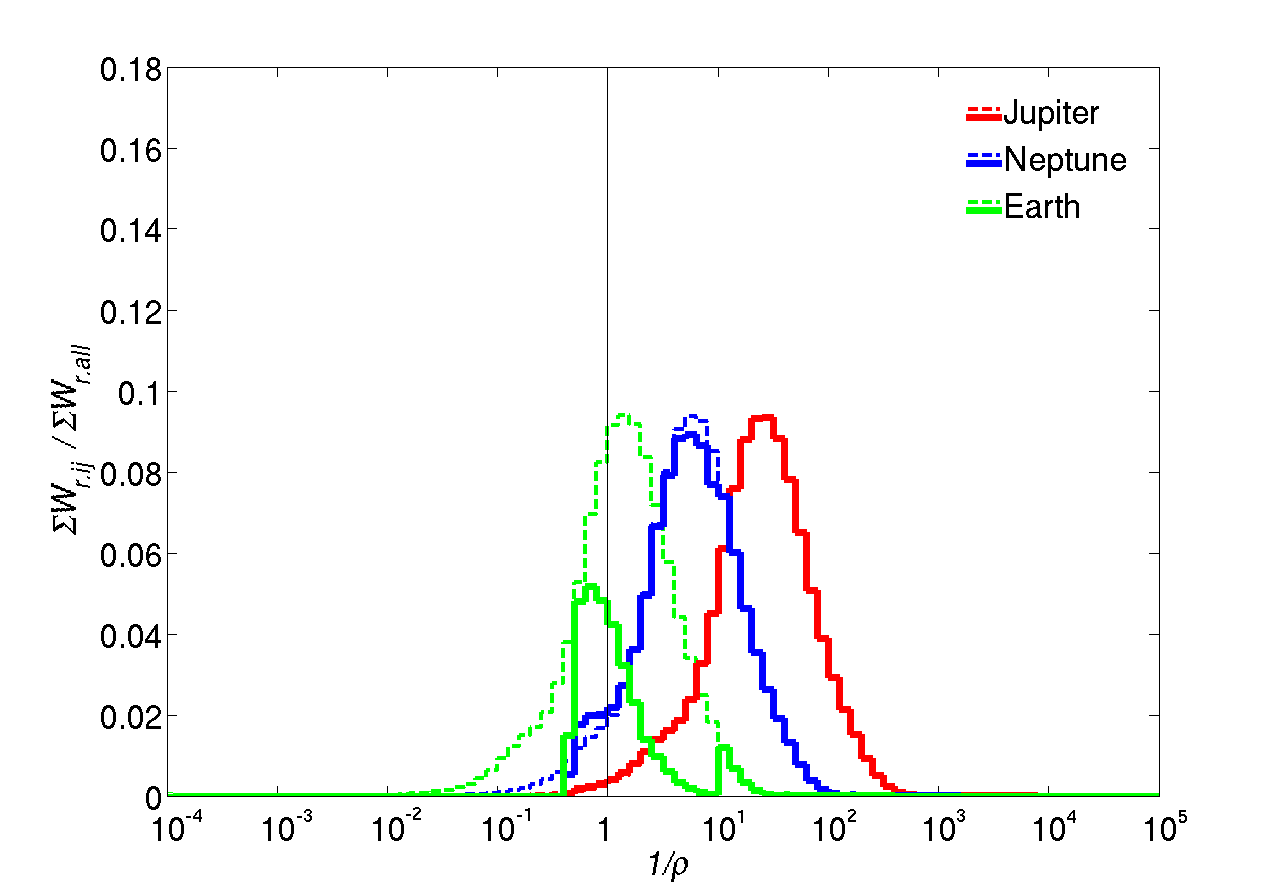}\\
\hspace*{-0.3in}
(c) Space-based $H$-band\\
\includegraphics[width=3.8in]{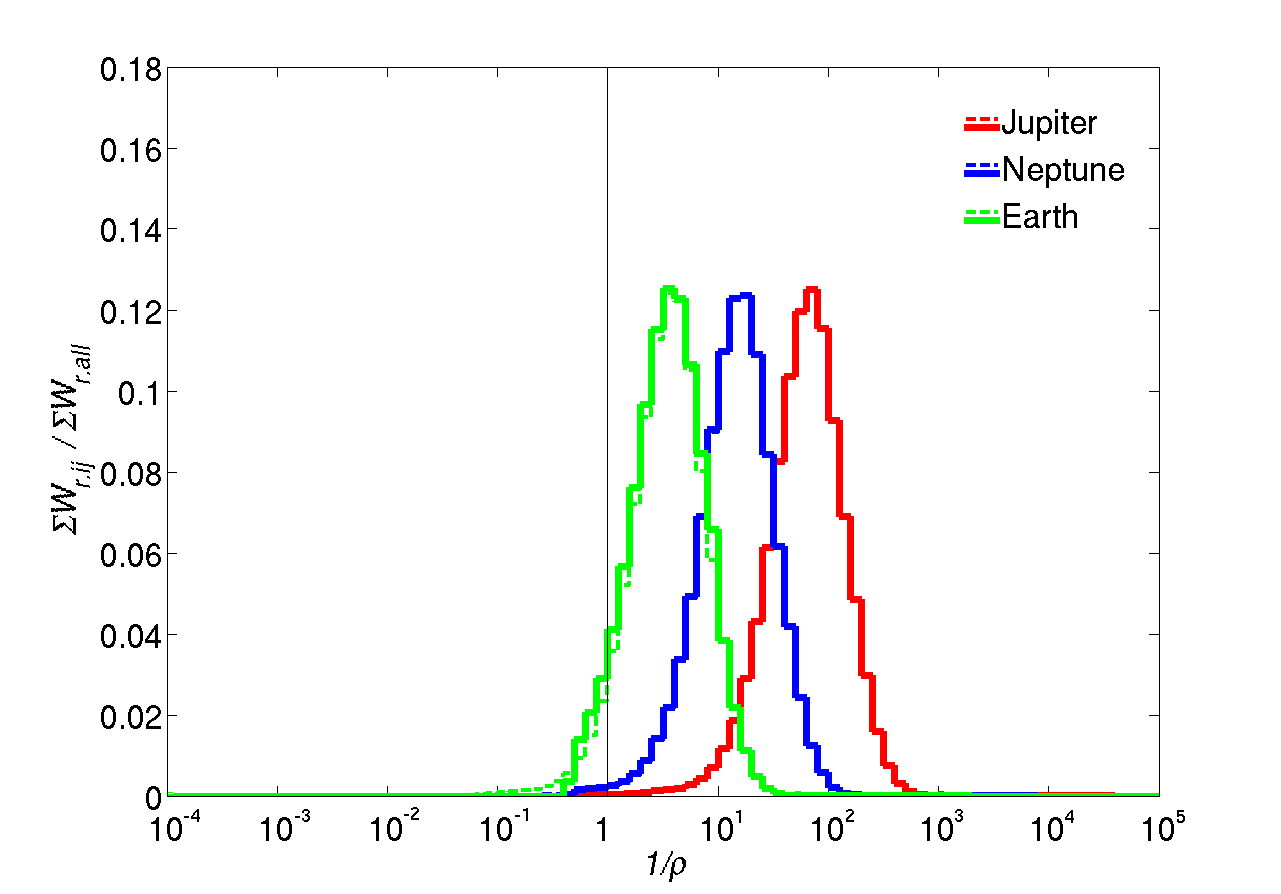}\\
}
\caption{Histogram of the angular Einstein radius $\theta_{\rm E}$ relative to the angular size of the source stars $\theta_{\rm S}$. Histograms are shown for ground-based $I$-band, ground-based $K$-band, and space-based $H$-band observations from top to bottom. Red, blue, and green histograms in each plot correspond to Jupiter-, Neptune-, and Earth-mass FFPs, respectively. The vertical line shows where $\theta_{\rm E}=\theta_{\rm S}$, indicating where finite source effects become dominant. The dashed histogram in each lens mass shows the distribution if a point source is assumed for all events, and the solid histograms are the distributions when the finite source effect is included.}
\label{fig:dist-map}
\end{figure}

\begin{figure*}
\hspace{-0.2in}
\begin{tabular}{cc}
(a) Ground-based $I$-band & (b) Ground-based $K$-band\\
\includegraphics[width=3.8in]{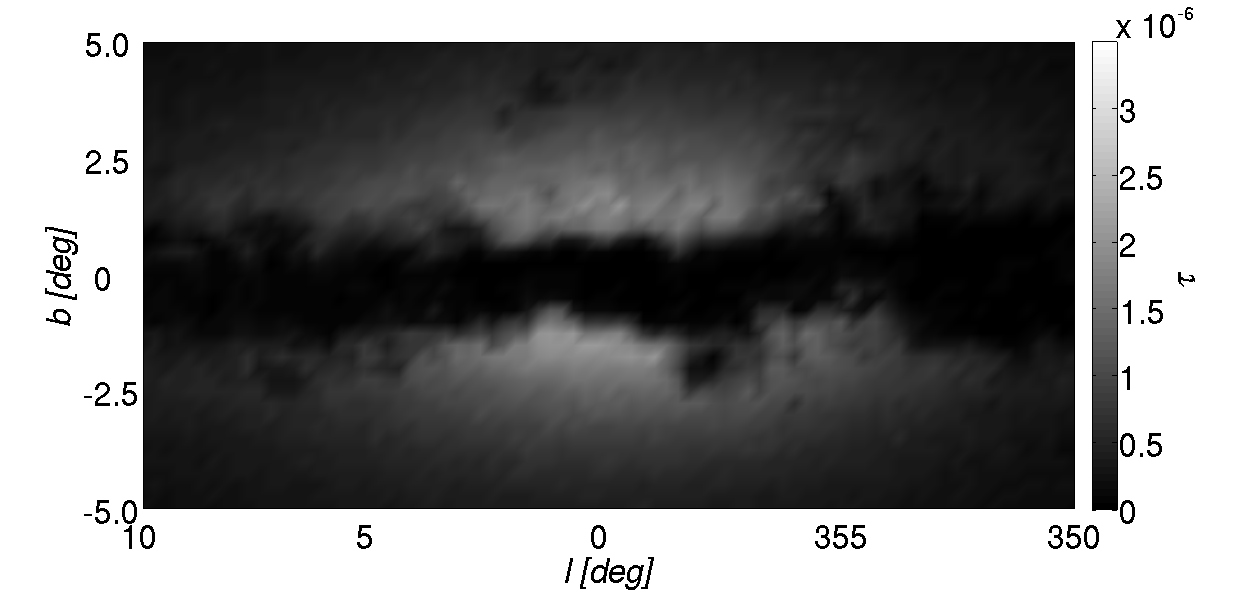} & \hspace{-0.3in}
\includegraphics[width=3.8in]{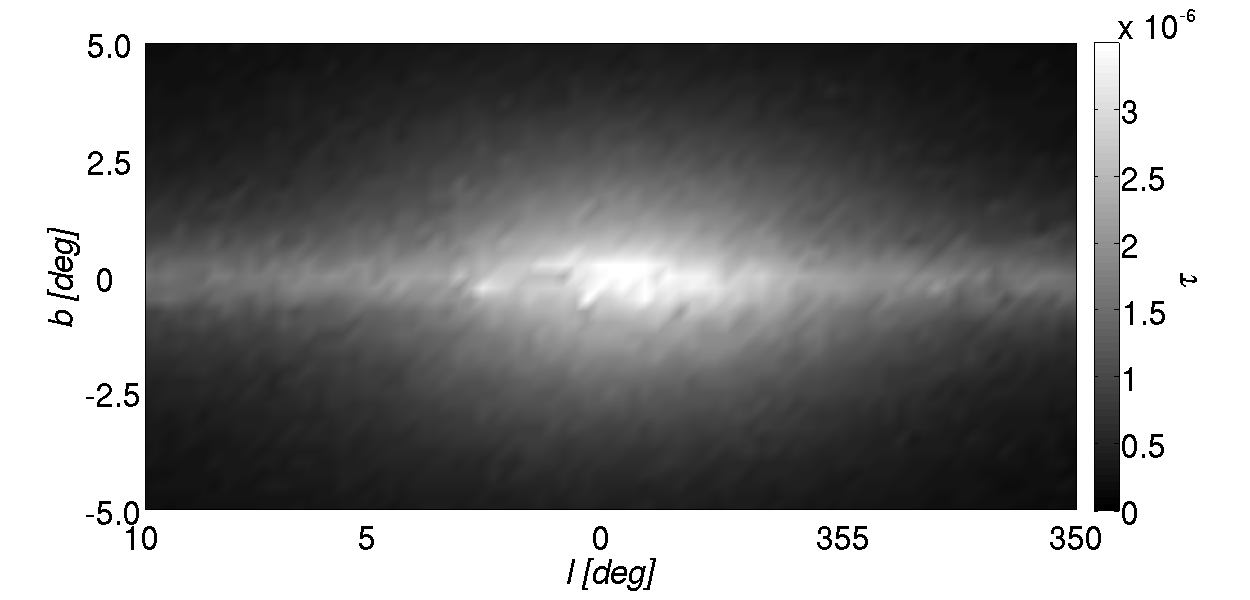} \\ 
(c) Space-based $H$-band & (d) VVV survey\\ 
\includegraphics[width=3.8in]{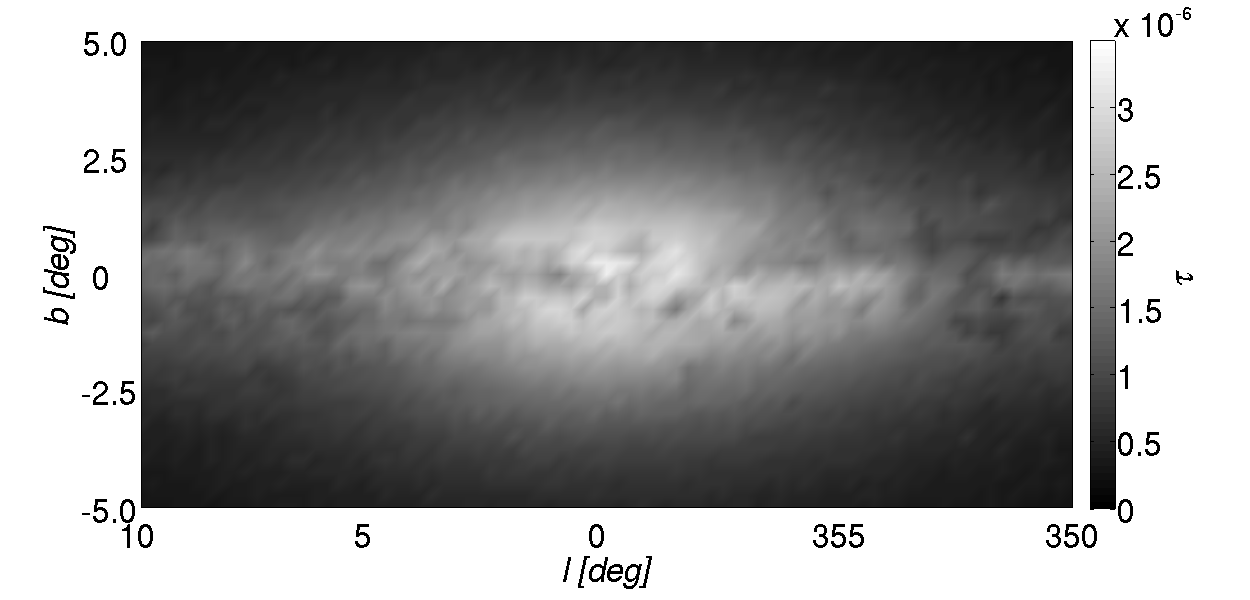} & \hspace{-0.2in}
\includegraphics[width=3.8in]{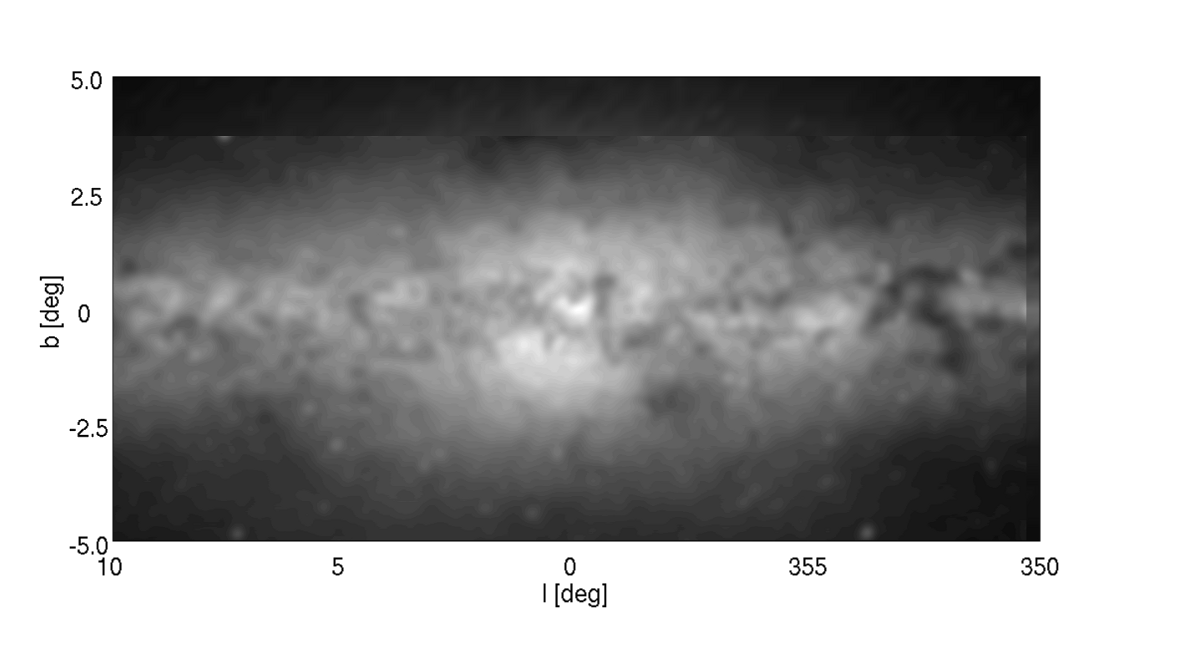} \\
\end{tabular}
\caption{Simulated maps of source-averaged stellar microlensing optical depth $\tau$.  {\em Top left:}\/ ground-based $I$-band view. {\em Top right:}\/ ground-based $K$-band view. {\em Bottom left:}\/ space-based $H$-band view. The colour bar scale is adjusted to the same range for comparison among bands. {\em Bottom right:}\/ a real $K_{\rm S}$-band mosaic of the inner Galaxy from the VVV survey. The image has been smoothed to render at approximately the same resolution as the microlensing maps and the intensity level has been squared, as the microlensing optical depth scales approximately as the square of the stellar density. VVV image credit: ESO/VVV Survey/D. Minniti, with acknowledgement to Ignacio Toledo and Martin Kornmesser.}
\label{fig:od-star-map}
\end{figure*}

Eq.\ref{eq:tau} allows for multiple star catalogues to be used for a given line of sight, provided the catalogues collectively  provide a contiguous non-overlapping sample over all relevant stellar parameters (magnitude, distance etc). In this case the inverse weight with source catalogue solid angle $\Omega_{{\rm cat},s}$ is required in order to appropriately average over sources. The solid angle sizes we use to generate stellar catalogues are given in Table \ref{tab:catals} and typically provide in excess of 2500 stars per catalogue for directions more crowded than Baade's window.
The use of catalogues of differing solid angles ensures the entire magnitude range is adequately sampled whilst keeping the overall simulation size to a minimum. 

It is important to generate catalogues containing all stars below the detection threshold as well as those above it, since the stars below the detection threshold can still act as lenses. We select stars with apparent $K$-band magnitudes in the range 0-99, split into four contiguous magnitude ranges, as detailed in Table \ref{tab:catals}. Since no cut is made in the $I-K$ or $H-K$ colour selection we can safely use the same catalogues for $I$-, $H$-, and $K$-band computations. The catalogues are used both for the lens catalogues $l$ and source catalogues $s$ so the number of computations scale as the square of the overall catalogue size. 

The catalogues are used to compute the microlensing optical depth, rate, and average duration using the formulae in Sect.\ref{subsec:microvals}. We compute these quantities for ordinary stellar microlenses using the masses generated by the \bes{} model. To compute microlensing quantities for FFPs we rerun the computation and replace each lens mass $M_i$ with a fixed FFP mass. This amounts to an assumption of an FFP population with an abundance of one FFP per Galactic star. We present calculations for three fixed FFP masses corresponding to Jupiter-, Neptune-, and Earth-mass FFPs. Because the size of the lensing cross-section depends on the lens mass, planetary mass lenses tend to provide finite source events more often as their Einstein radius can be comparable to the angular size of the microlensed source star. Figure \ref{fig:dist-map} shows the histograms of the Einstein radius $\theta_{\rm E}$ (relative to source size $\theta_{\rm S}$) towards Baade's window ($l, b$)=($1^{\circ}, -4^{\circ}$). The histograms are constructed by weighting each lens-source pair by the rate weight defined by Eq.\ref{eq:wrate}. Each plot shows three histograms for Jupiter-mass (red), Neptune-mass (blue), and Earth-mass (green) FFPs for the point source regime only (dashed) and including finite source effects (solid). For the ground-based surveys ($I$ and $K$), the Jupiter-mass FFP distribution indicates that most of the events are within the point source regime. The Neptune-mass FFP distribution includes more finite source events than Jupiter-mass FFPs, and the $1/\rho < 1$ region shows the benefit of magnification boost for events where $1/\rho < 1$ (see Figure \ref{fig:amp}) as the solid curve is drawn over the dashed curve. For the Earth-mass FFPs, the finite source regime becomes dominant, and the event likelihood dramatically decreases from the massive FFP cases. The secondary peak at $log_{10}(1/\rho < 1)=1$ is caused by the population of faint stars in bulge and our signal-to-noise limit (S/N>50). The space-based $H$ band shows a smooth distribution compared to the ground-based simulations; the point source regime is dominant for all lens masses and the space-based sensitivity is capable of observing low-mass FFPs for our assumed detectability limits.

To check our computation accuracy the optical depth and average duration are calculated for ten independent subsamples drawn from the catalogues. Typically the microlensing calculations involve summing ${\cal O} (10^8)$ lens-source pairs for every line of sight more crowded than Baade's window. Therefore, in practice the agreement between these subsamples is good, with a standard deviation of around $2\%$ in the average timescale and less than $1\%$ in the optical depth.

The finite source effect is included by using a pre-computed look-up table of magnification versus source radius and impact parameter, shown plotted in Figure \ref{fig:amp}. Two curves illustrate the boundary of minimum magnification of 1.34 (corresponding to $u_{\rm max} = 1$ for the point source regime) and 1.02 ($u_{\rm max} = 3$).

\section{Results} \label{sec:res}	

The aim of this paper is to present detailed microlensing maps using a consistent stellar population synthesis model for FFP populations, accounting for signal-to-noise, the unresolved stellar background and finite source size effects. We compare the FFP and stellar maps directly to assess possible spatial systematics between the FFP and stellar microlensing distributions caused by differences in finite source size effects. Maps are presented for three survey configurations: ground-based $I$-band (OGLE/MOA-like sensitivity), ground-based $K$-band (VISTA-like sensitivity) and space-based $H$-band (\euclid{}-like sensitivity).

\subsection{Stellar lens simulation} \label{subsec:star}
\subsubsection{Optical depth} 	\label{subsubsec:od-star}

Figure~\ref{fig:od-star-map} shows maps of stellar microlensing optical depth for ground-based $I$-band, ground-based $K$-band, and space-based $H$-band observations. The $I$-band map shows clearly the strong effects of extinction towards the Galactic plane and closely resembles the distribution seen in the $I$-band microlensing map presented in \cite{ker09} for an earlier version of the \bes{} model (version 06-05), which contained a more massive single-component bulge. 

The most noticeable difference is that the optical depth distribution maximises in Figure \ref{fig:od-star-map} at a much lower value of around $3.5\times 10^{-6}$ compared to $5\times 10^{-6}$ for the ``resolved source'' map in Fig. 1 of \cite{ker09}. There are two main reasons for the difference. First, the bulge model employed in the 06-05 \bes{} model had a much higher total mass of $2.05\times 10^{10}$ M$_{\odot}$ compared to a total mass of roughly half of this amount for the two-component bulge model used here. Second, the maps in the present paper are signal-to-noise limited rather than magnitude limited and take explicit account of the background light from unresolved stars, a factor that was not included by \cite{ker09}.

For spaced-based microlensing with a maximum impact parameter $u_{\rm max}=1$, the maximum optical depth for the space-based $H$-band survey becomes $\tau \simeq 3.2\times10^{-6}$, which is a little less than the maximum optical depth for the ground-based $K$-band survey. This is because the average line-of-sight source distance is smaller as a near-IR survey is sensitive to more numerous nearby low-mass stars. If we impose $u_{\rm max}=3$ for spaced-based microlensing, then the optical depth becomes almost an order of magnitude larger due to the $u_{\rm max}^2$ dependence. We note, however, that for signal-to-noise limited calculations the optical depth does not scale exactly as $u_{\rm max}^2$ since detectability depends on $u_{\rm t}$ which, for fainter sources, becomes sensitive to finite source size effects.

\cite{sumi13} analysed 474 microlensing events observed towards the Galactic bulge, determining an optical depth towards ($l < 5^{\circ}, <b> = -2.^{\circ}26$) of $\tau_{200} = 3.49^{+0.81}_{-0.66}\times10^{-6}$ in the $I$ band for events with durations below 200 days. A recent re-analysis by \cite{sumi16} has revised this downwards to $\tau_{200} = 2.61^{+0.61}_{-0.49}\times10^{-6}$ for the same region. Our result at  ($l = 1^{\circ}, b = -2.^{\circ}25$) in the $I$ band is $\tau = 1.64\times10^{-6}$ which is only around $63\%$ of the revised value, but discrepant from it by less than $2 \, \sigma$. Other factors which can contribute a difference include differences between the MOA-II filter throughput and the Johnson-Cousins $I$-band filter used for our simulation. \cite{awip16} show that the \bes{} model, whilst underestimating the inner bulge optical depth, provides a good fit to the observed optical depth for $|b| > 3^{\circ}$ and is in reasonable agreement with the MOA observed timescale distribution over pretty much all of the survey area. For this reason a simple global scaling factor of 1.6 can be applied to match the revised MOA observations.

\subsubsection{Einstein timescale} \label{subsubsec:te-star}

\begin{figure}
{\centering
\hspace*{-0.2in}
(a) Ground-based $I$-band.\\ \hspace*{-0.2in}
\includegraphics[width=3.8in]{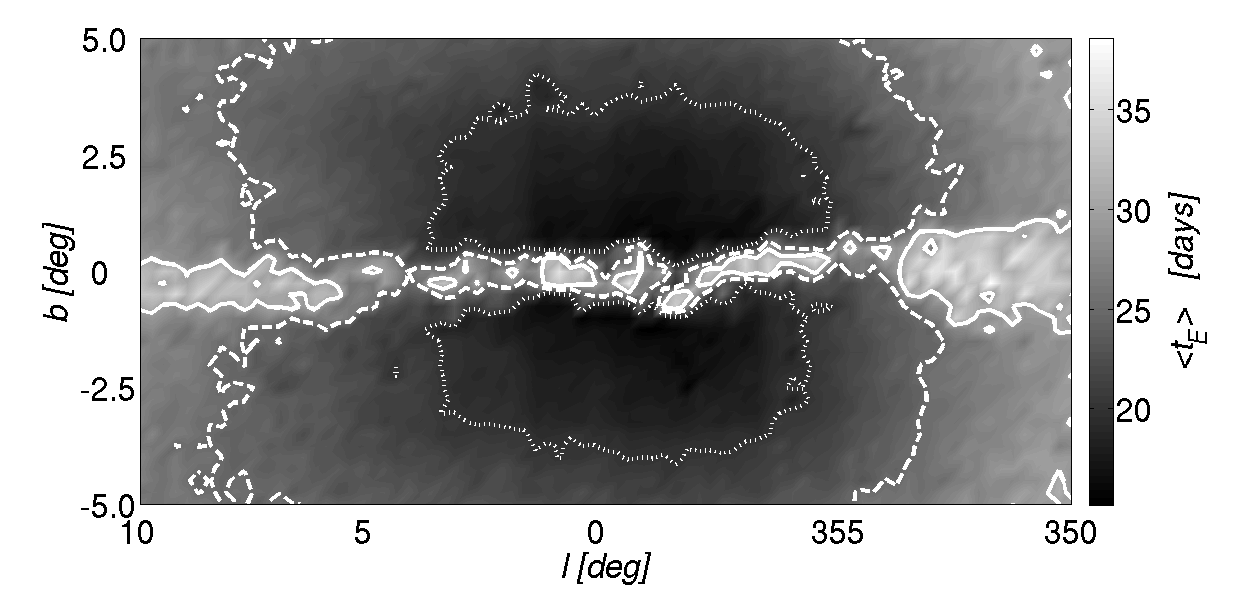}\\
\hspace*{-0.2in}
(b) Ground-based $K$-band.\\ \hspace*{-0.2in}
\includegraphics[width=3.8in]{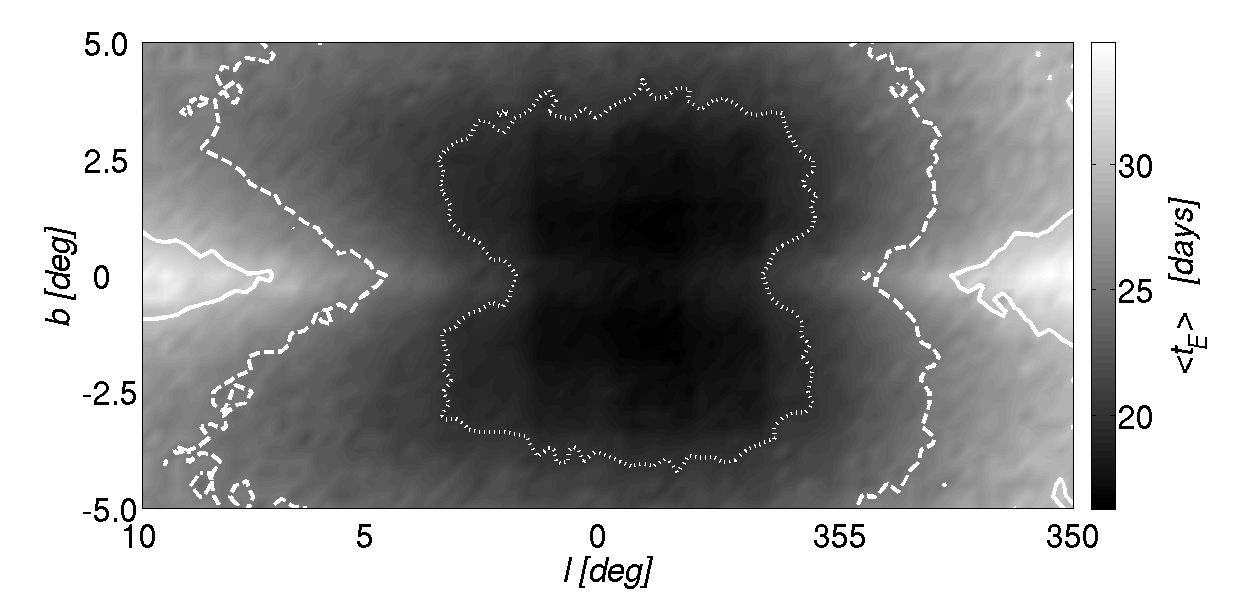}\\
\hspace*{-0.2in}
(c) Space-based $H$-band.\\ \hspace*{-0.2in}
\includegraphics[width=3.8in]{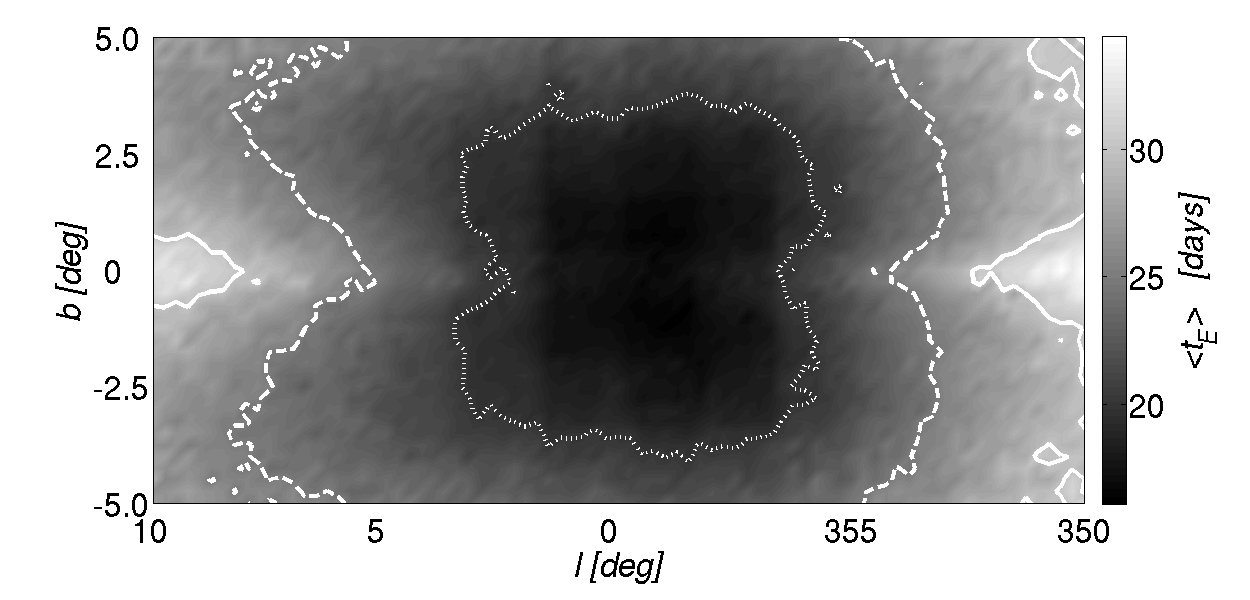}\\
}
\caption{Maps of average Einstein timescale $\langle t_E \rangle [days]$: ground-based $I$-band, ground-based $K$-band, and space-based $H$-band observations from top to bottom. The dotted, dashed, and solid contours are 20, 25, and 30 days, respectively.}
\label{fig:te-star-map}
\end{figure}

Maps of the average Einstein timescale $\langle t_{\rm E} \rangle$ are presented in Figure \ref{fig:te-star-map}. For all bands, the distribution pattern is almost symmetric about $b=0$, with the average event duration decreasing going towards the Galactic Centre. The long timescales near the Galactic plane ($b\simeq0$) for ground-based $I$-band corresponds to the low optical depth region in Figure \ref{fig:od-star-map} where there is essentially no microlensing signal in the $I$ band. Away from the mid-plane all photometric bands show typical timescales of around 18 days at low Galactic latitudes rising to around 25 days at higher latitudes near the edge of our simulation. The shape of the timescale contours differs significantly from those in \cite{ker09} though the bulge model used here differs substantially, as does the event selection procedure (Kerins, Robin and Marshall did not compute S/N-limited maps). The spatial distribution of event timescales presented here appears to be in much better agreement with the observed distribution by the OGLE-III survey \citep{wyrzykowski15} as well as with the MOA survey \citep{sumi13,awip16}. We note that, aside from extinction effects visible close to the mid-plane in $I$-band, the maps for all three configurations show a very high degree of correspondence indicating that observations across optical and near-IR passbands are observing very similar populations in terms of their line-of-sight integrated spatial and kinematic distributions.

\subsubsection{Event rate} 

\begin{figure}	
{\centering		
\hspace*{-0.2in}
(a) Ground-based $I$-band.\\ \hspace*{-0.2in}
\includegraphics[width=3.8in]{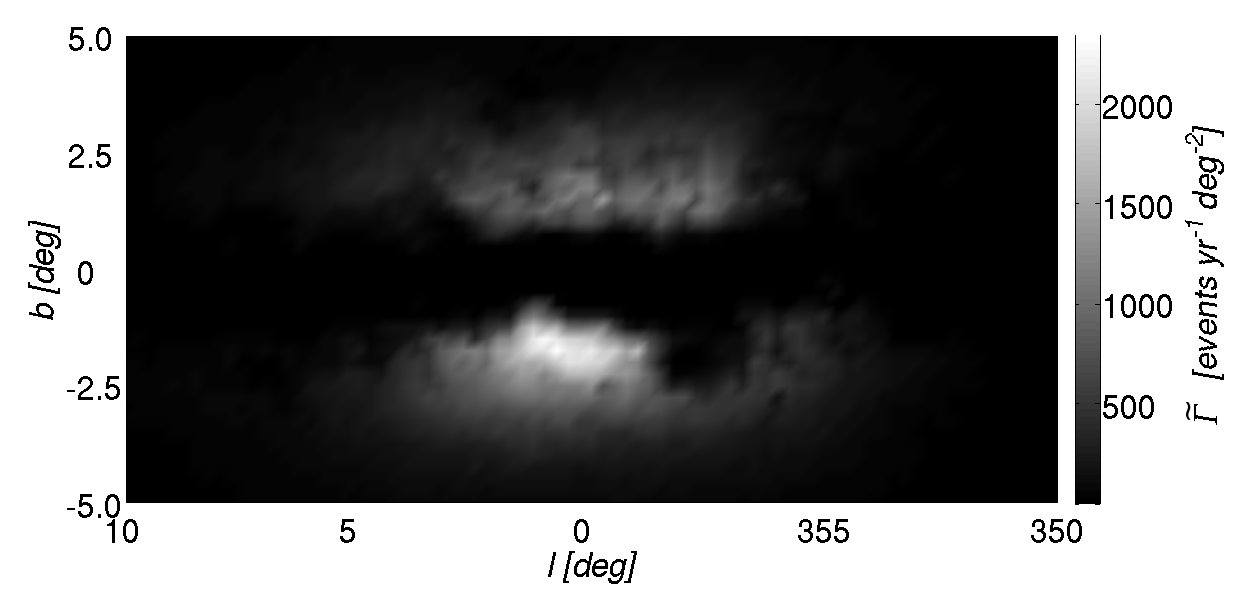}\\
\hspace*{-0.2in}
(b) Ground-based $K$-band.\\ \hspace*{-0.2in}
\includegraphics[width=3.8in]{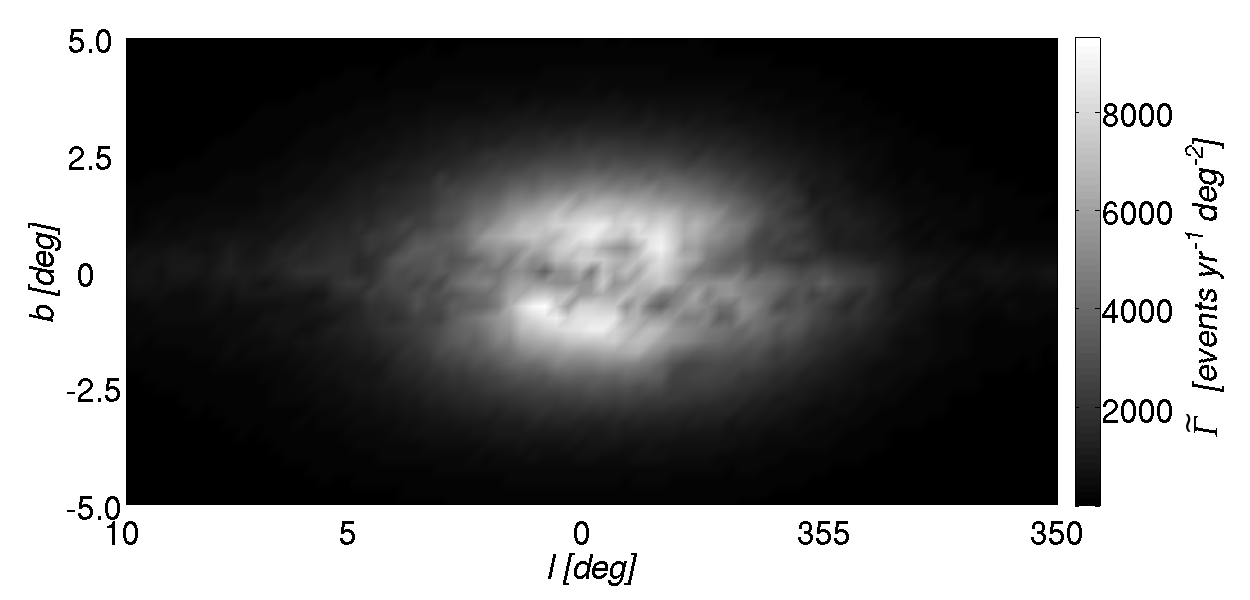}\\
\hspace*{-0.2in}
(c) Space-based $H$-band.\\ \hspace*{-0.2in}
\includegraphics[width=3.8in]{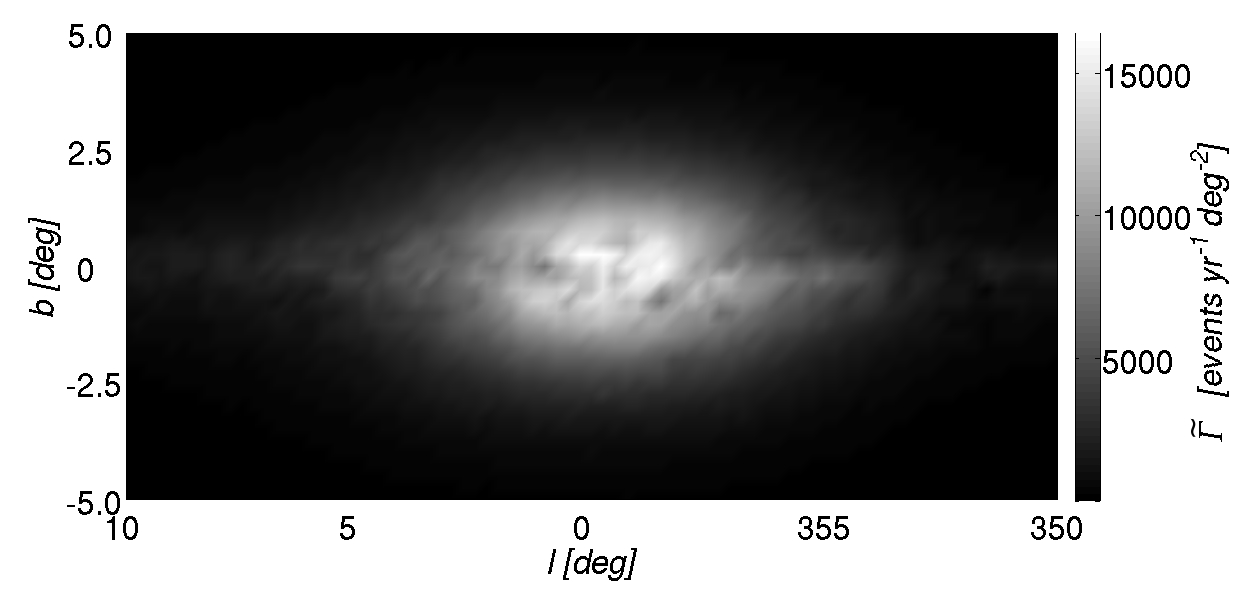}\\
}
\caption{Maps of the event rate $\Gamma\times N_*$ [events yr$^{-1}$ deg$^{-2}$]: ground-based $I$-band, ground-based $K$-band, and space-based $H$-band observations from top to bottom. The values shown in the colour bars are raw values which are not multiplied by the 1.6 correction factor discussed in Section \ref{subsubsec:od-star}.}
\label{fig:er-star-map}
\end{figure}

Figure \ref{fig:er-star-map} presents maps of the stellar event rate per unit sky area. Unlike the timescale map the event rate distribution is sensitive to the observational setup. We caution that we do not include any timescale threshold, nor do we account for finite time sampling or sources of photometric inefficiency (e.g. the impact of source saturation). Our rate predictions therefore represent a firm theoretical maximum rate which assumes a survey with perfect sampling and photometric efficiency. Since we are concerned in this paper primarily with the relative rate between different survey strategies and with the relative spatial signatures of FFP events versus ordinary microlensing events we do not need to worry here about this limitation of approach.

The ``hot-spot'' of the event rate is at essentially the same location for all simulated ground-based surveys, but for our space-based $H$-band simulation the hot-spot is located closer to the Galactic Centre. The location of the predicted hot spot region for ground-based surveys agrees well with the findings of \cite{sumi13}. In terms of the magnitude of the rate, over a target area of 3.2 deg$^2$ centred at ($l, b$) = ($0^{\circ}.97, -2^{\circ}.26$) \cite{sumi16} determined a rate $\Gamma = 3.41_{-0.34}^{+0.38}\times10^{-5}$ star$^{-1}$ year$^{-1}$ in $I$-band. Our result for this region is $\Gamma = 3.1\times10^{-5}$ star$^{-1}$ year$^{-1}$ after the correction factor of 1.6 is applied (see Section \ref{subsubsec:od-star}), which is consistent with the revised MOA value.  

\begin{table*}
 \centering
\begin{minipage}{5.0in}
  \caption{The maximuml values and area integrated FFP event rates in SN mode for $100\%$ detection efficiency and assuming $u_{\rm max} = 1$. The maximum rates are in units of events yr$^{-1}$ deg$^{-2}$ with the rate contrast in square brackets for FFPs. The hot-spot location is given in the Galactic coordinates $(l, b)$. The global rates are in units of events yr$^{-1}$ which covers the whole simulation area of 200 deg$^2$. The errors of these maximum rates are all $<$1\%, and a correction factor of 1.6 has been applied to the rate.}
  \label{tab:maxer-ffp}
  \begin{tabular}{lllll}
  \hline
  Lens mass & & $I$-band (Ground) & $K$-band (Ground) & $H$-band (Space) \\ \hline \hline
          & max & 3,700  & 15,000  & 26,000  \\ 
  Stellar & location & $(0.^{\circ}75, -1.^{\circ}75)$ & $(1^{\circ}, -0.^{\circ}75)$ & $(0.^{\circ}25, 0.^{\circ}25)$ \\
          & global & 58,000 & 450,000 & 710,000 \\ \hline
          & max & 200 [-0.3\%]  & 850 [0.4\%]  & 1,500 [1.4\%]  \\ 
  Jupiter & location & $(0.^{\circ}75, -1.^{\circ}75)$ & $(1^{\circ}, -0.^{\circ}75)$ & $(-1.^{\circ}50, 0^{\circ})$ \\
          & global & 3,100 & 25,000 & 40,000 \\ \hline
          & max & 36 [1.6\%]  & 170 [-6.5\%]   & 340 [0.05\%]  \\ 
  Neptune & location & $(0.^{\circ}75, -1.^{\circ}75)$ & $(1^{\circ}, -0.^{\circ}75)$ & $(-1.^{\circ}50, 0^{\circ})$ \\
          & global & 460 & 4,900 & 8,900 \\ \hline
          & max & 1.1 [-22\%]   & 12 [-20\%]  & 68 [-4.9\%]  \\ 
  Earth   & location & $(0.^{\circ}25, -2^{\circ})$ & $(1^{\circ}, -1.^{\circ}50)$ & $(-1.^{\circ}50, 0.^{\circ}50)$ \\
          & global & 18 & 350 & 2,000 \\ \hline
\end{tabular}
\end{minipage}
\end{table*}

\subsection{FFP lens simulation}  \label{subsec:ffp}	

Having calibrated the ordinary stellar microlensing yields we performed FFP simulations for three FFP masses: Jupiter-mass, Neptune-mass, and Earth-mass. The simulations were performed using the same catalogues generated for the stellar microlensing simulations, but with the stellar masses replaced by the FFP mass. All other properties such as the lens distance, source characteristics and relative proper motion remain the same. This equates to an assumed FFP abundance of 1 per Galactic star.

\subsubsection{FFP timescale} 

\begin{landscape}
\begin{figure}
{\centering
\vspace*{1.0in} \hspace{-0.3in}
\begin{tabular}{ccc}
Ground-based $I$-band, Jupiter-sized FFP & Ground-based $I$-band, Neptune-sized FFP & Ground-based $I$-band, Earth-sized FFP\\
\includegraphics[width=3.4in]{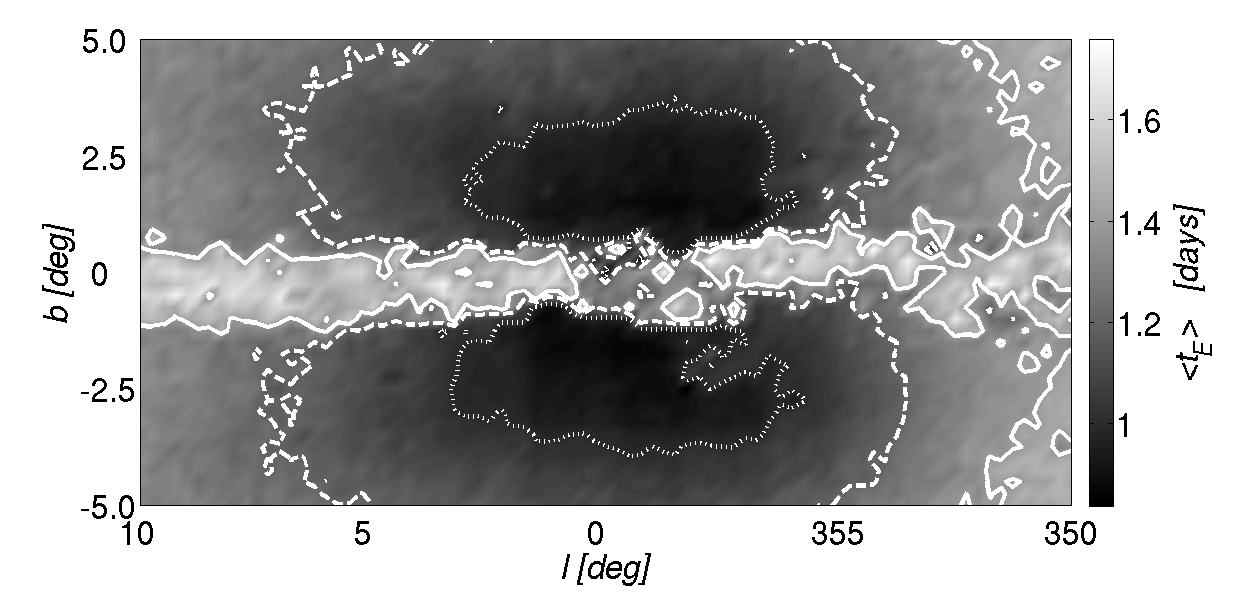} & \hspace{-0.3in}
\includegraphics[width=3.4in]{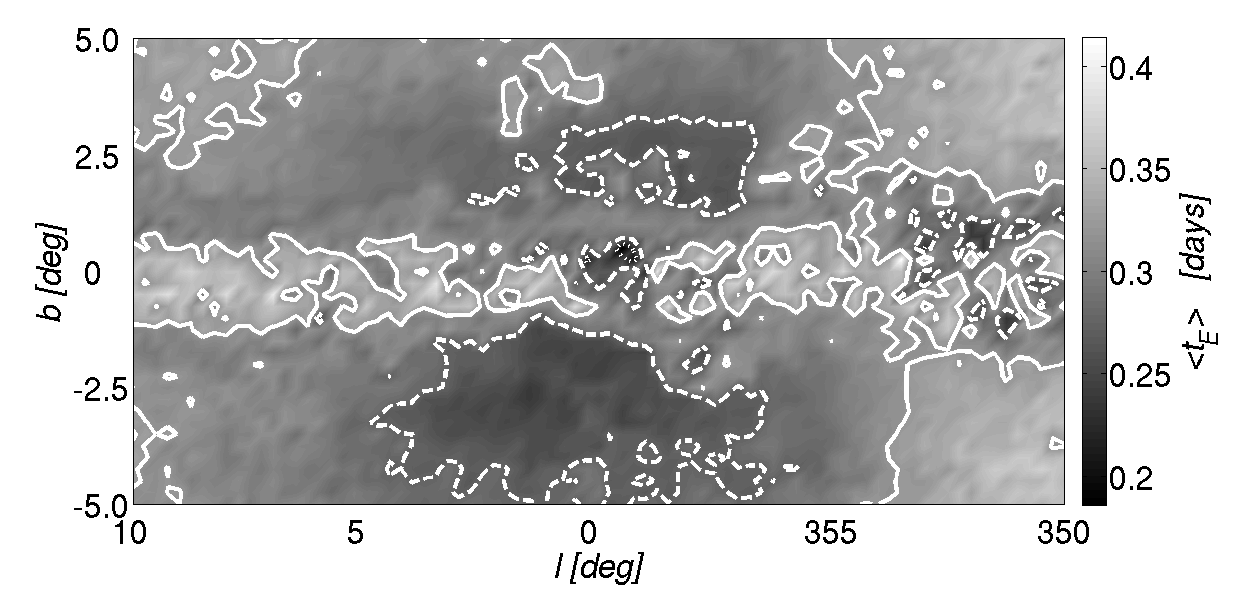} & \hspace{-0.3in}
\includegraphics[width=3.4in]{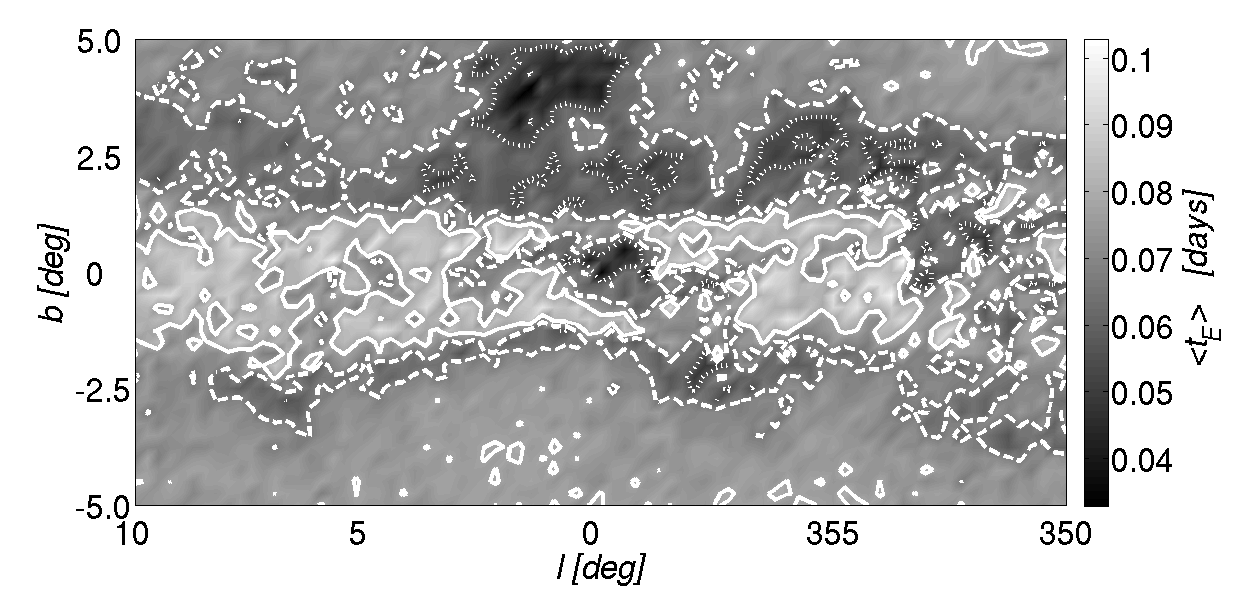} \\
Ground-based $K$-band, Jupiter-sized FFP & Ground-based $K$-band, Neptune-sized FFP & Ground-based $K$-band, Earth-sized FFP\\
\includegraphics[width=3.4in]{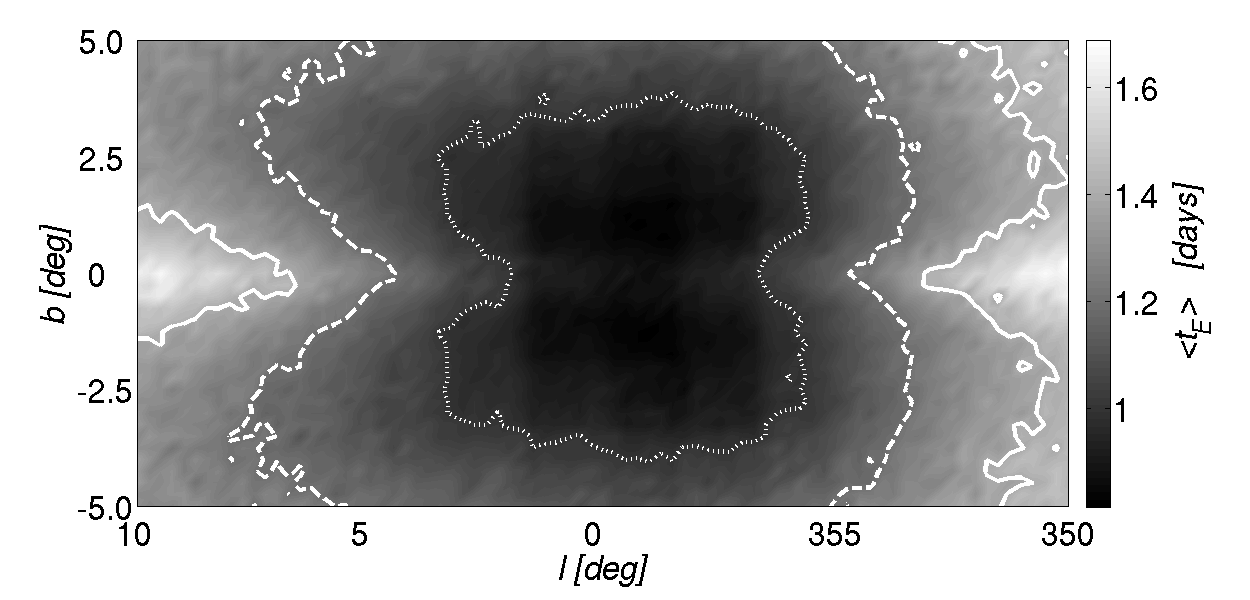} & \hspace{-0.3in}
\includegraphics[width=3.4in]{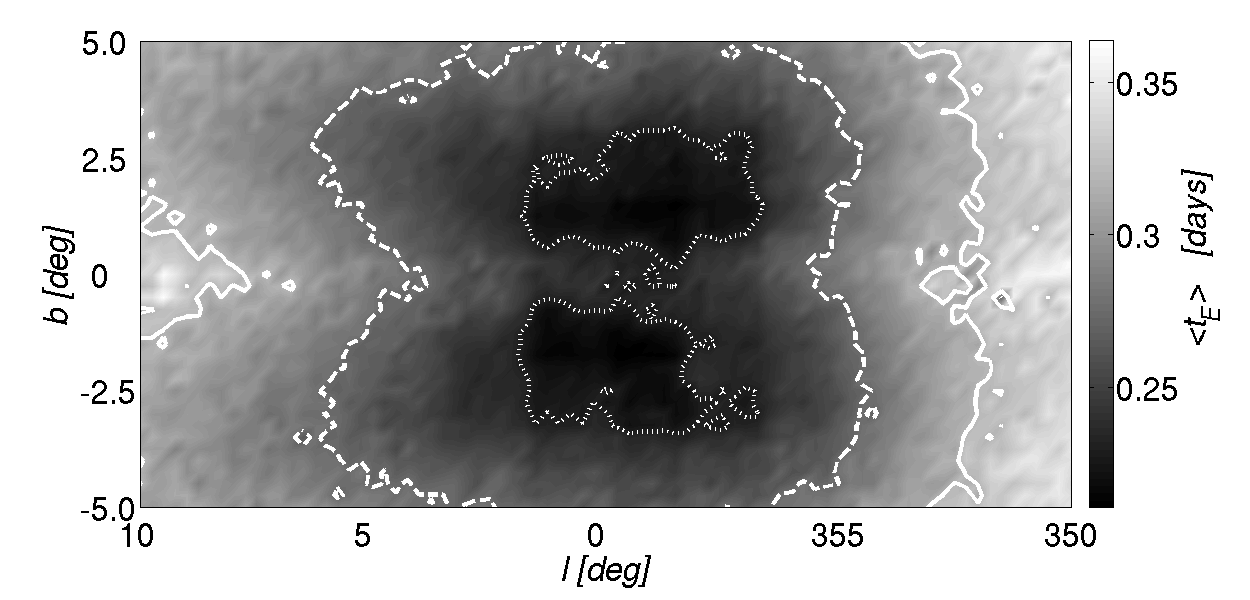} & \hspace{-0.3in}
\includegraphics[width=3.4in]{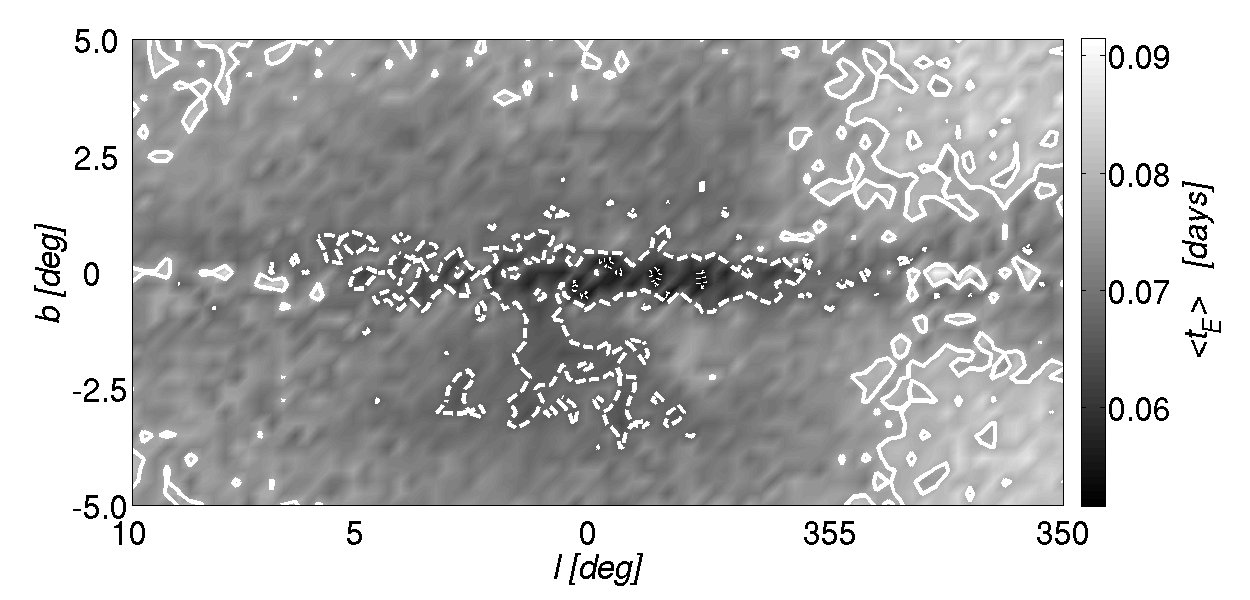} \\
Space-based $H$-band, Jupiter-sized FFP & Space-based $H$-band, Neptune-sized FFP & Space-based $H$-band, Earth-sized FFP\\
\includegraphics[width=3.4in]{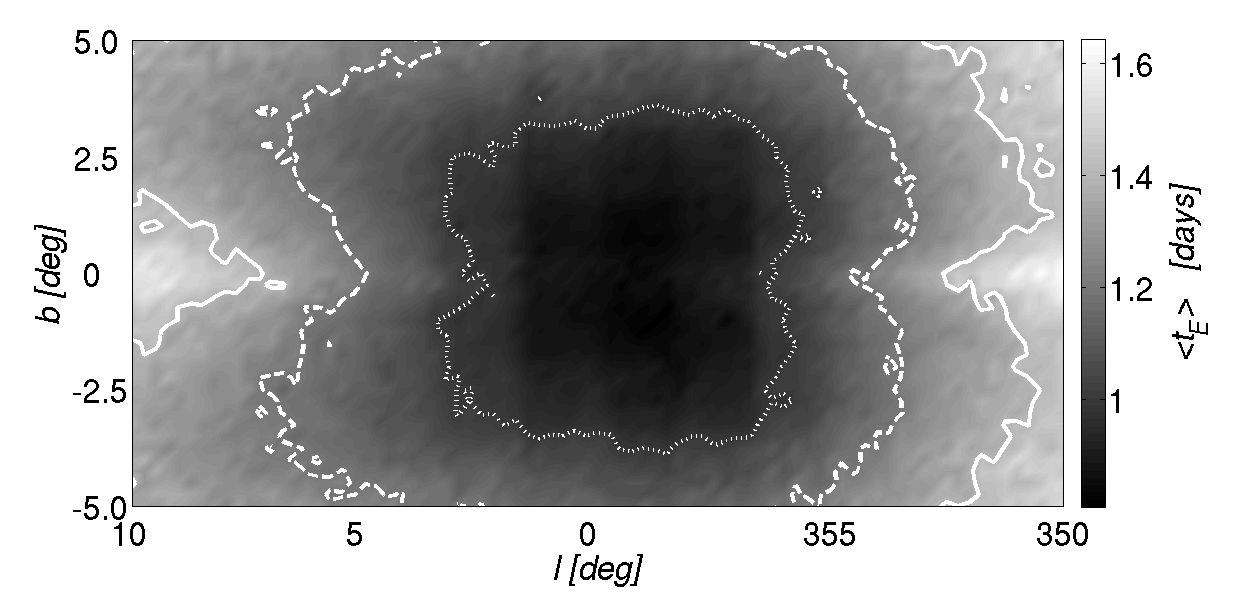} & \hspace{-0.3in}
\includegraphics[width=3.4in]{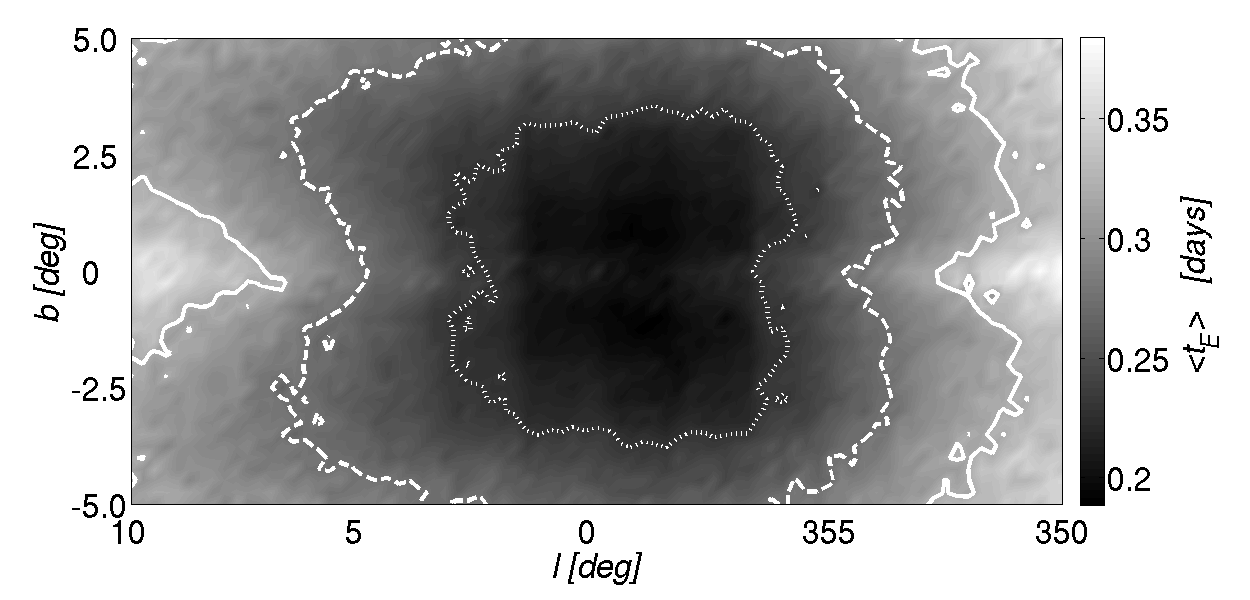} & \hspace{-0.3in}
\includegraphics[width=3.4in]{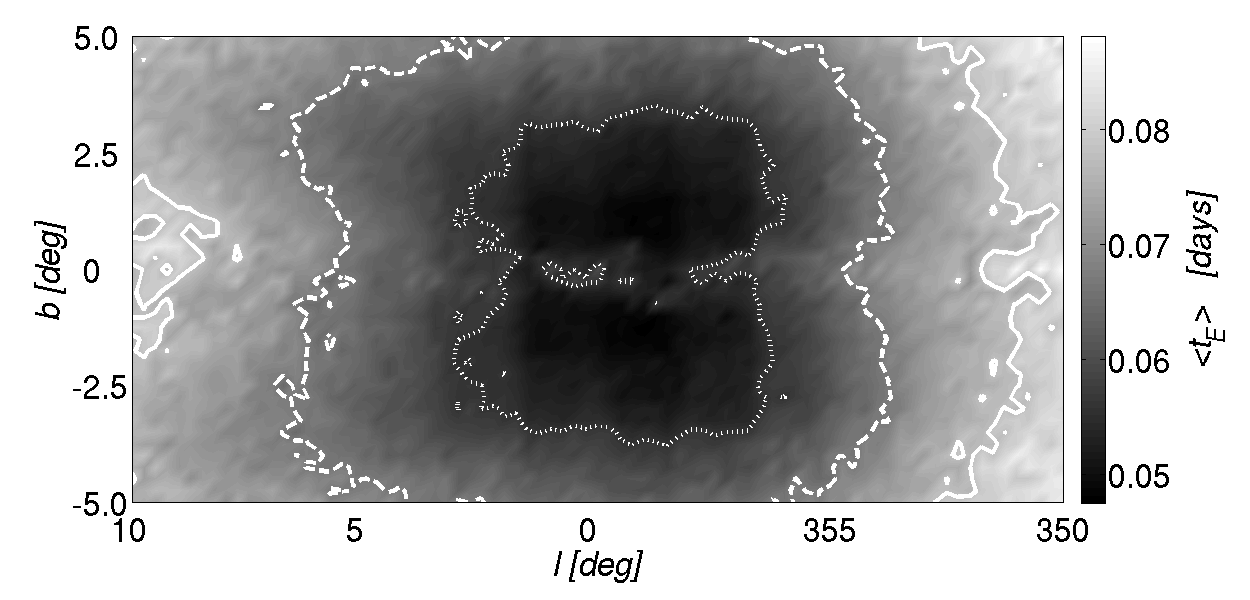} \\
\end{tabular}
}
\caption{Maps of average Einstein timescale $<t_E> [days]$. The columns are Jupiter-mass, Neptune-mass, and Earth-mass FFPs from left to right. The rows are ground-based $I$-band, ground-based $K$-band, and space-based $H$-band observations from top to bottom. The dotted, dashed, and solid lines are set at 1.0, 1.2 and 1.4 $(M_{\rm FFP}/M_J)^{1/2}$ days.}
\label{fig:te-ffp-map}
\end{figure}
\end{landscape}

Figure \ref{fig:te-ffp-map} shows the spatial variation in the FFP average timescale. The contours levels indicate durations of 1.0, 1.2 and 1.4 $(M_{\textup{FFP}}/M_{\textup{Jup}})^{1/2}$ days. 

The ground-based $I$-band maps exhibit the longest timescales closest to the plane (towards $b=0$), though the optical depth is negligible in this region so that only rare nearby disk lens-disk source microlensing can be detected. This is the same tendency as the $I$-band maps of stellar lens masses seen in Section \ref{subsubsec:te-star}. The maps of space-based $H$-band show an almost FFP-mass independent structure whilst the ground-based maps show some evolution in the contours with FFP mass. As we mentioned in the event distribution (Figure \ref{fig:dist-map}), the noisy map for Earth-mass FFPs with ground-based surveys reflects the low rate signal and therefore the low likelihood of any event detection. For Neptune-mass FFPs the timescale maps of $I$-band and $K$-band observations show a strong difference not just because of the lessened effect of dust in the $K$ band, but also because finite source size microlensing is more significant for the $I$-band simulations, as highlighted in Figure~\ref{fig:dist-map}. Thus, the near-IR is a better choice to observe FFPs down to Neptune mass from the ground. 

\subsubsection{FFP event rate} 

\begin{landscape}
\begin{figure}
{\centering
\vspace*{1.0in} \hspace{-0.3in}
\begin{tabular}{ccc}
Ground-based $I$-band, Jupiter-sized FFP & Ground-based $I$-band, Neptune-sized FFP & Ground-based $I$-band, Earth-sized FFP\\
\includegraphics[width=3.4in]{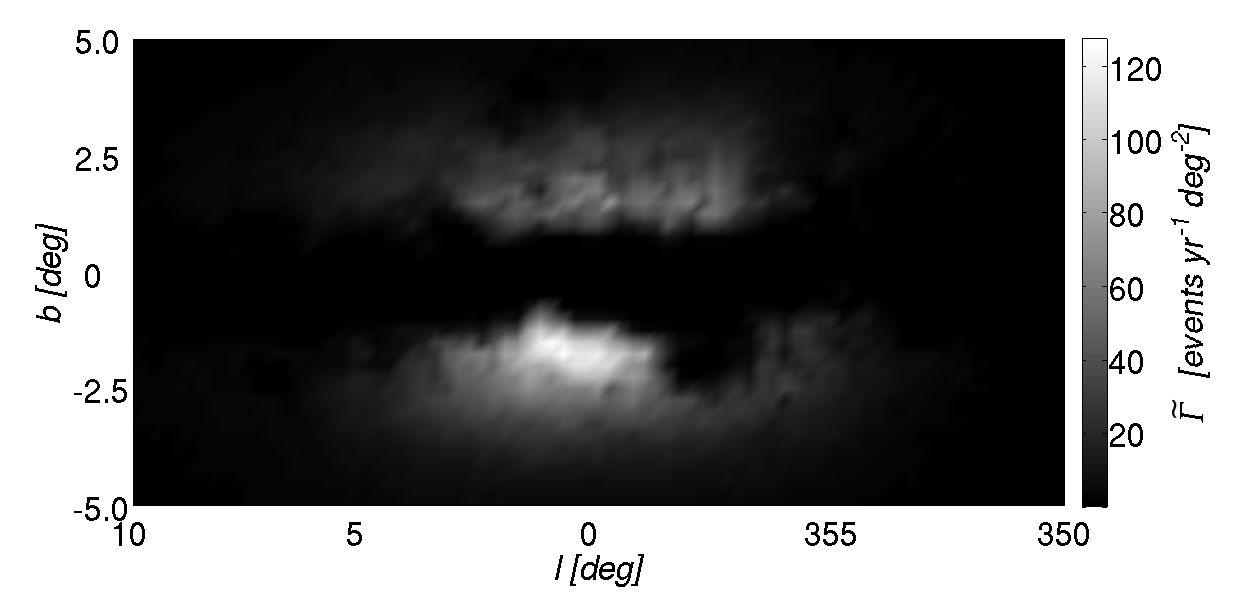} & \hspace{-0.3in}
\includegraphics[width=3.4in]{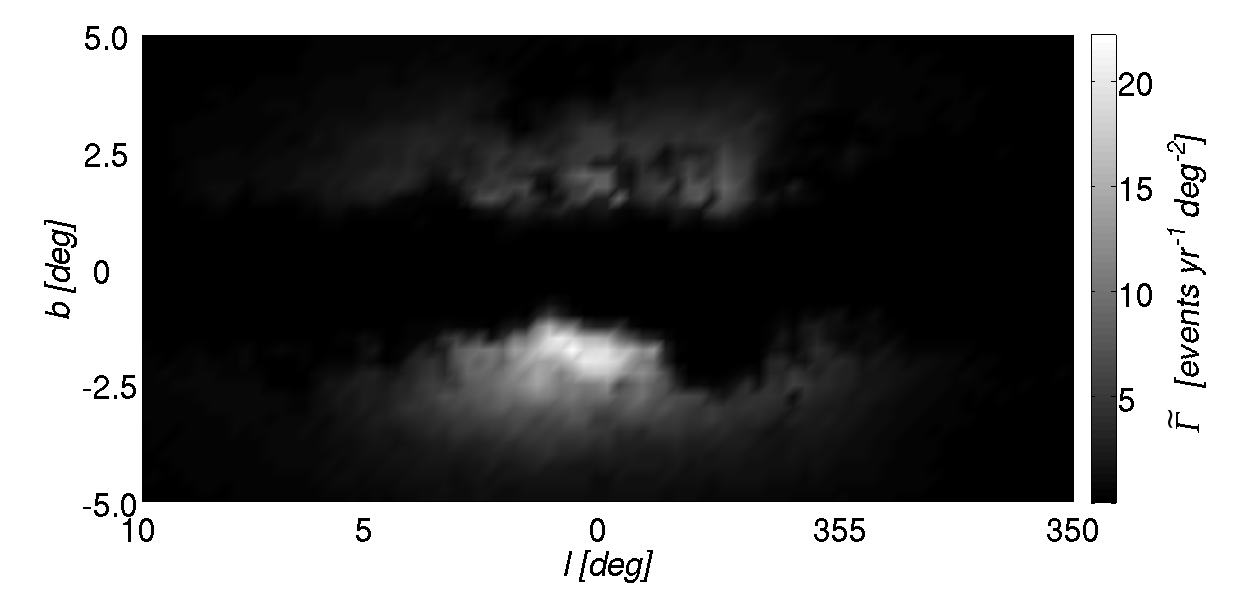} & \hspace{-0.3in}
\includegraphics[width=3.4in]{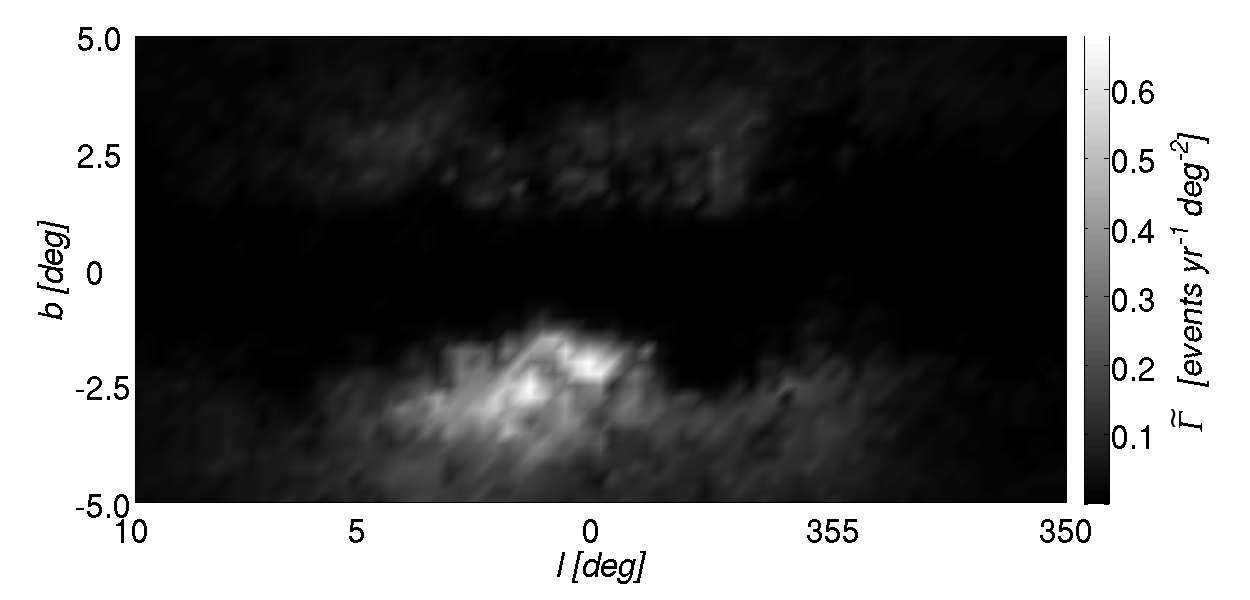} \\
Ground-based $K$-band, Jupiter-sized FFP & Ground-based $K$-band, Neptune-sized FFP & Ground-based $K$-band, Earth-sized FFP\\
\includegraphics[width=3.4in]{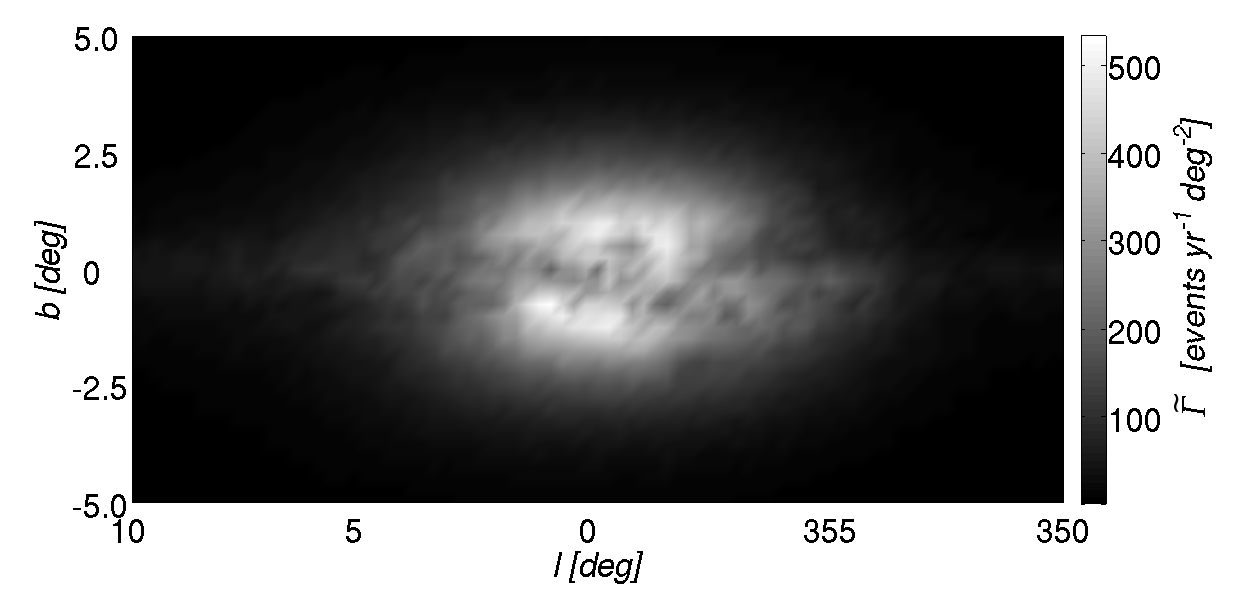} & \hspace{-0.3in}
\includegraphics[width=3.4in]{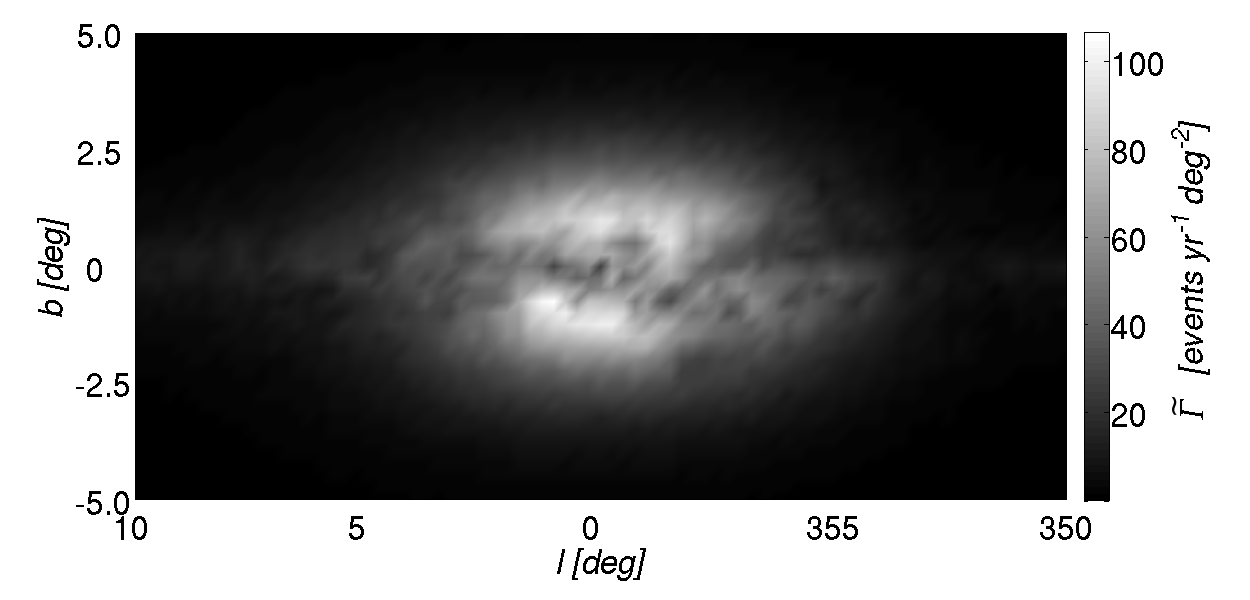} & \hspace{-0.3in}
\includegraphics[width=3.4in]{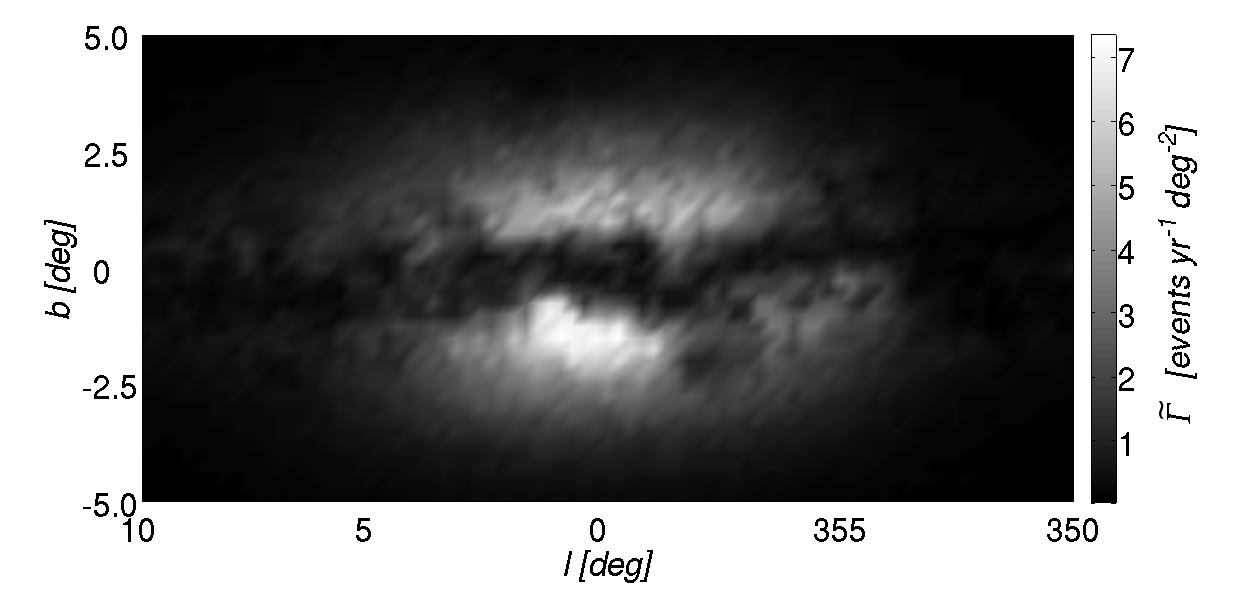} \\
Space-based $H$-band, Jupiter-sized FFP & Space-based $H$-band, Neptune-sized FFP & Space-based $H$-band, Earth-sized FFP\\
\includegraphics[width=3.4in]{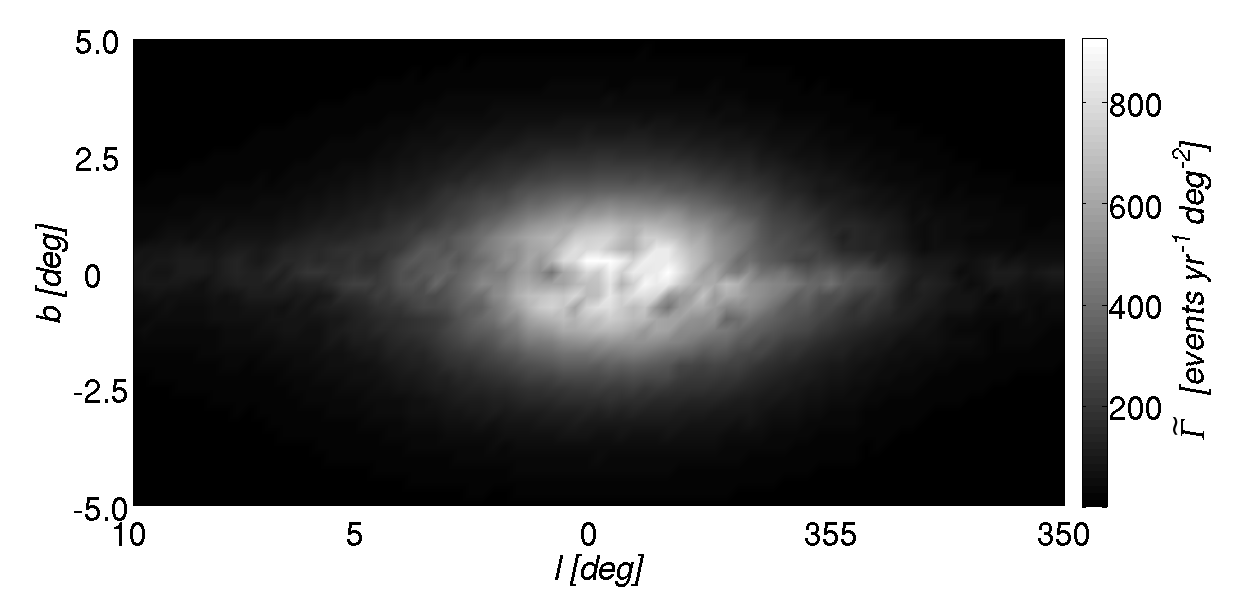} & \hspace{-0.3in}
\includegraphics[width=3.4in]{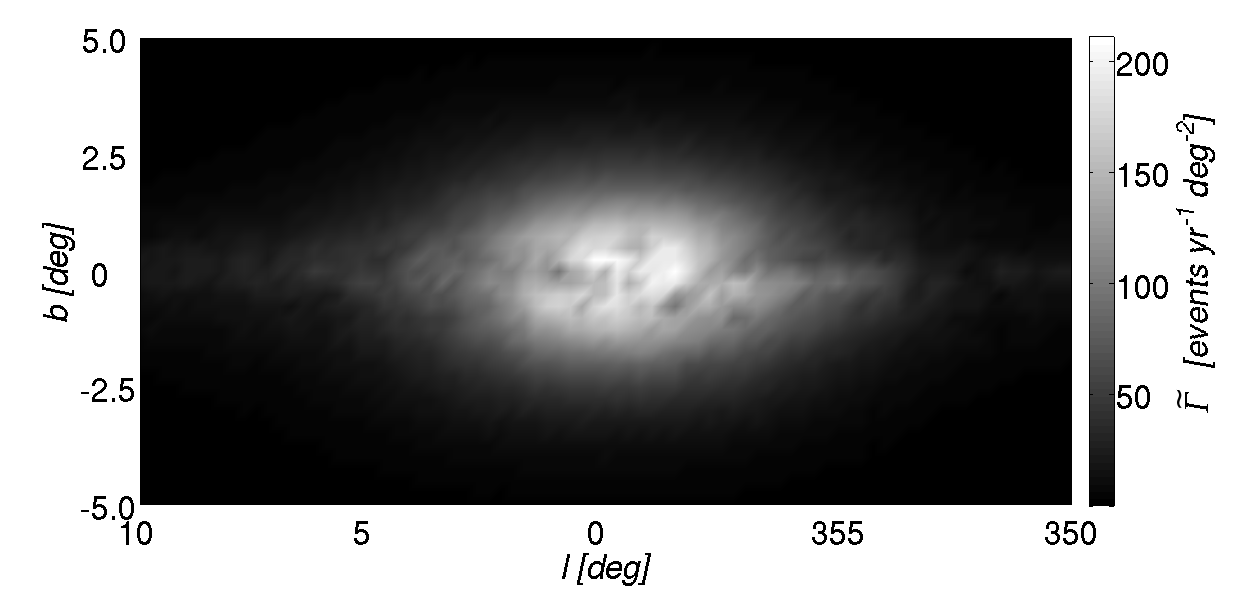} & \hspace{-0.3in}
\includegraphics[width=3.4in]{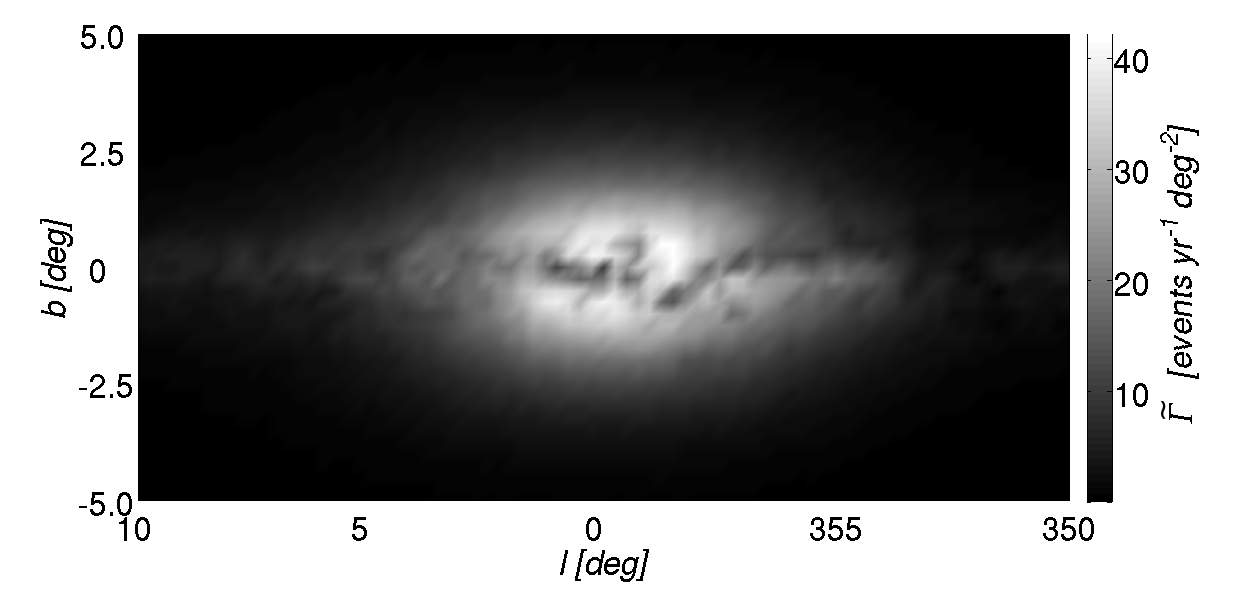} \\
\end{tabular}
}
\caption{Maps of the event rate $\Gamma\times N_*$ in units of events yr$^{-1}$.deg$^{-2}$. The columns are Jupiter-mass, Neptune-mass, and Earth-mass FFPs from left to right. The rows are ground-based $I$-band, ground-based $K$-band, and space-based $H$-band observations from top to bottom. The values shown in the colour bars are not multiplied by a correction factor of 1.6.}
\label{fig:er-ffp-map}
\end{figure}
\end{landscape}

Figure~\ref{fig:er-ffp-map} shows the maps of event rate per unit sky area for FFP lenses with rows and columns arranged according to survey and FFP lens mass. Through all bands and lens masses the event rate hot-spot coincides with that for stellar lenses, as observed by \cite{sumi13}. For all bands the overall event rate is highest for Jupiter mass FFPs but declines at each simulated mass scale by a factor of 4 or more. 

\begin{landscape}
\begin{figure}
{\centering
\vspace*{1.0in} \hspace{-0.3in}
\begin{tabular}{ccc}
Ground-based $I$-band, Jupiter-sized FFP & Ground-based $I$-band, Neptune-sized FFP & Ground-based $I$-band, Earth-sized FFP\\
\includegraphics[width=3.4in]{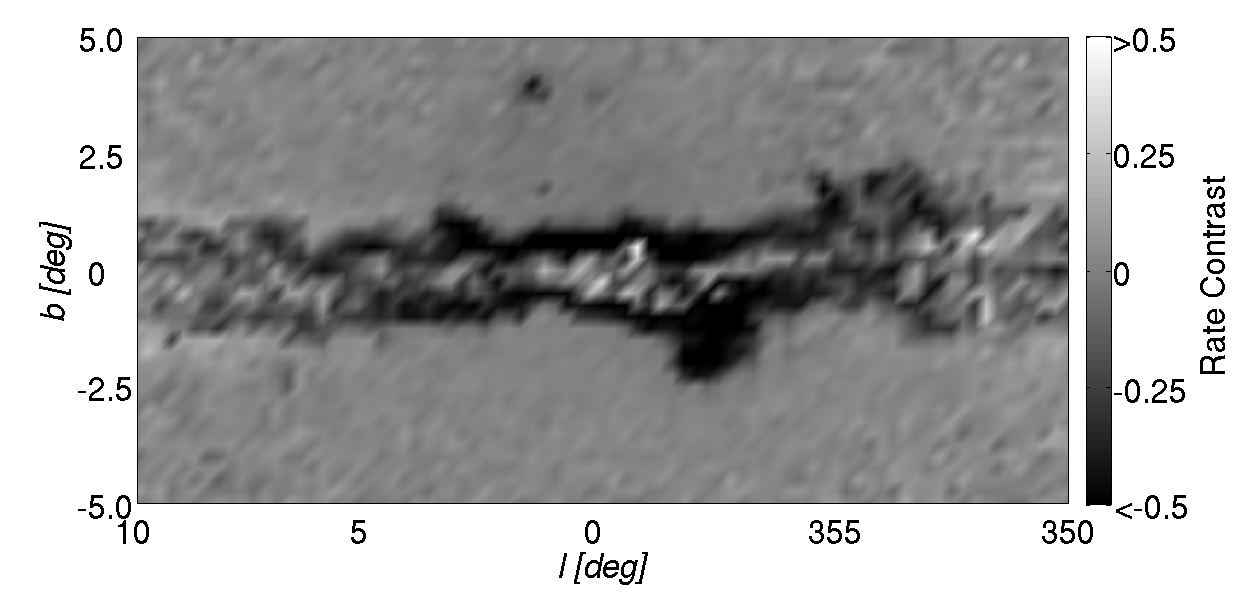} & \hspace{-0.3in}
\includegraphics[width=3.4in]{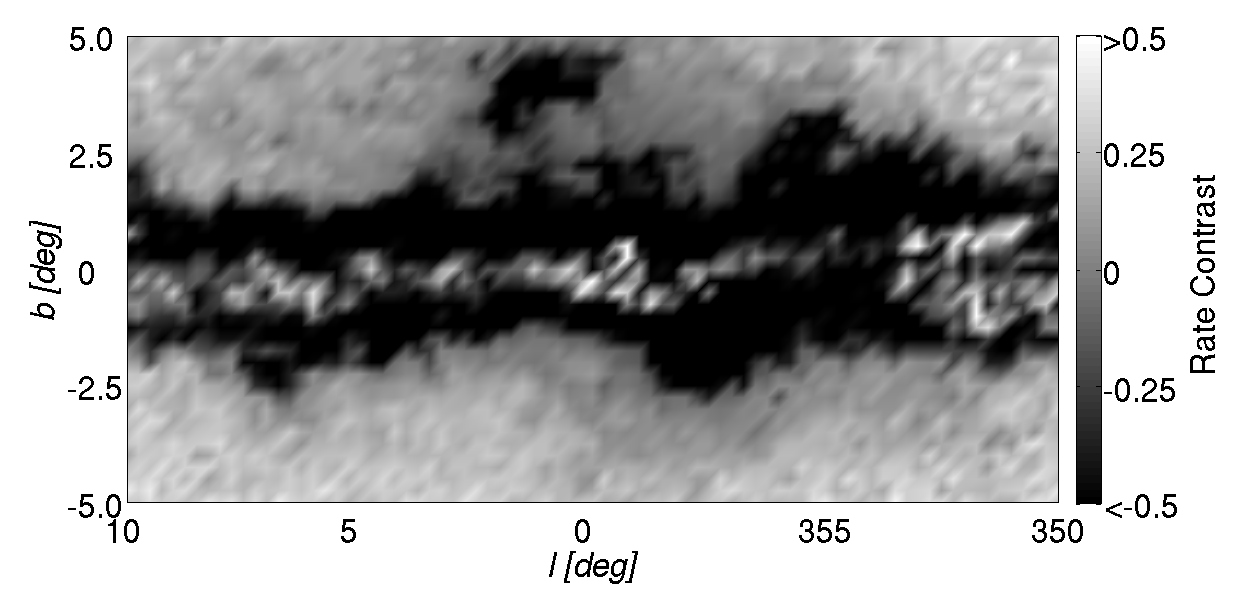} & \hspace{-0.3in}
\includegraphics[width=3.4in]{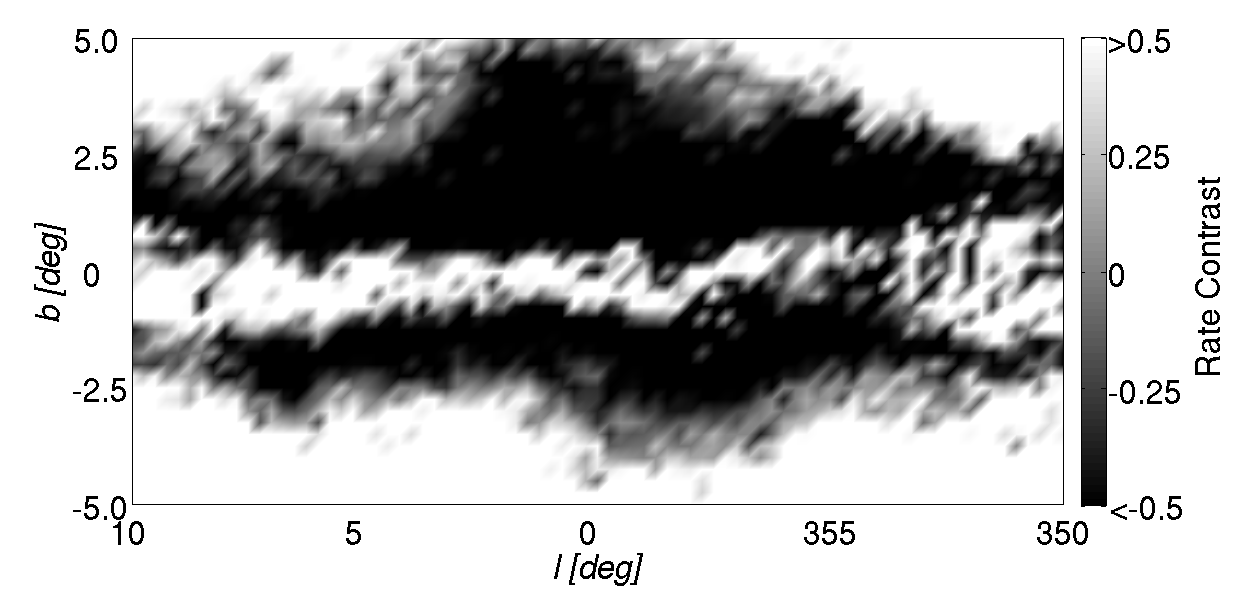} \\
Ground-based $K$-band, Jupiter-sized FFP & Ground-based $K$-band, Neptune-sized FFP & Ground-based $K$-band, Earth-sized FFP\\
\includegraphics[width=3.4in]{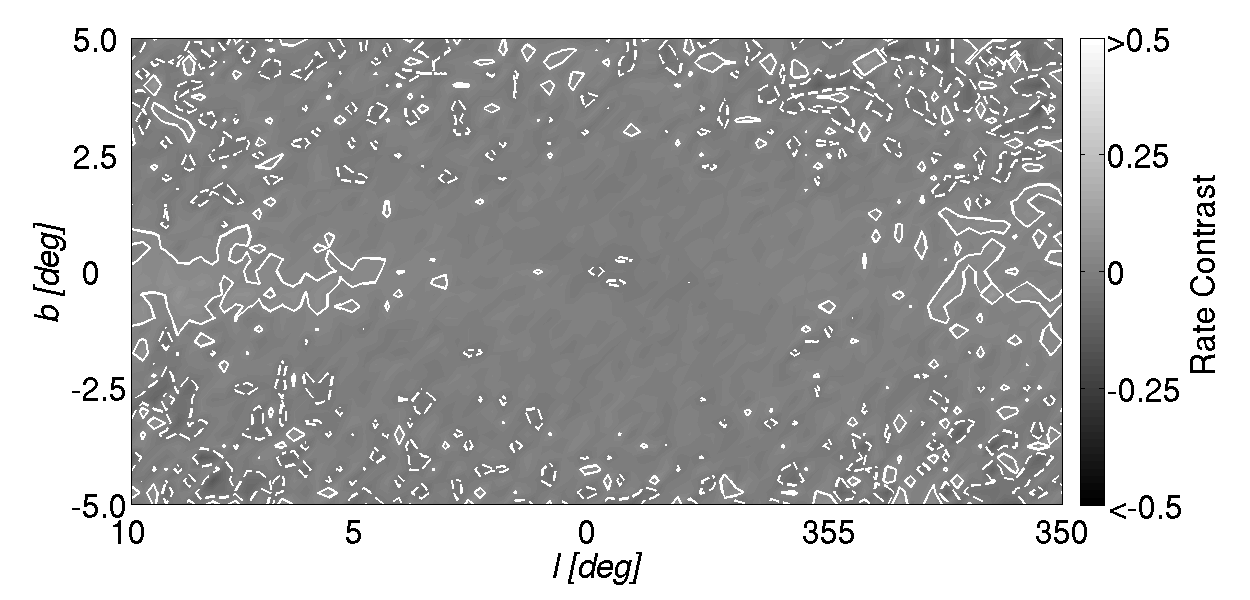} & \hspace{-0.3in}
\includegraphics[width=3.4in]{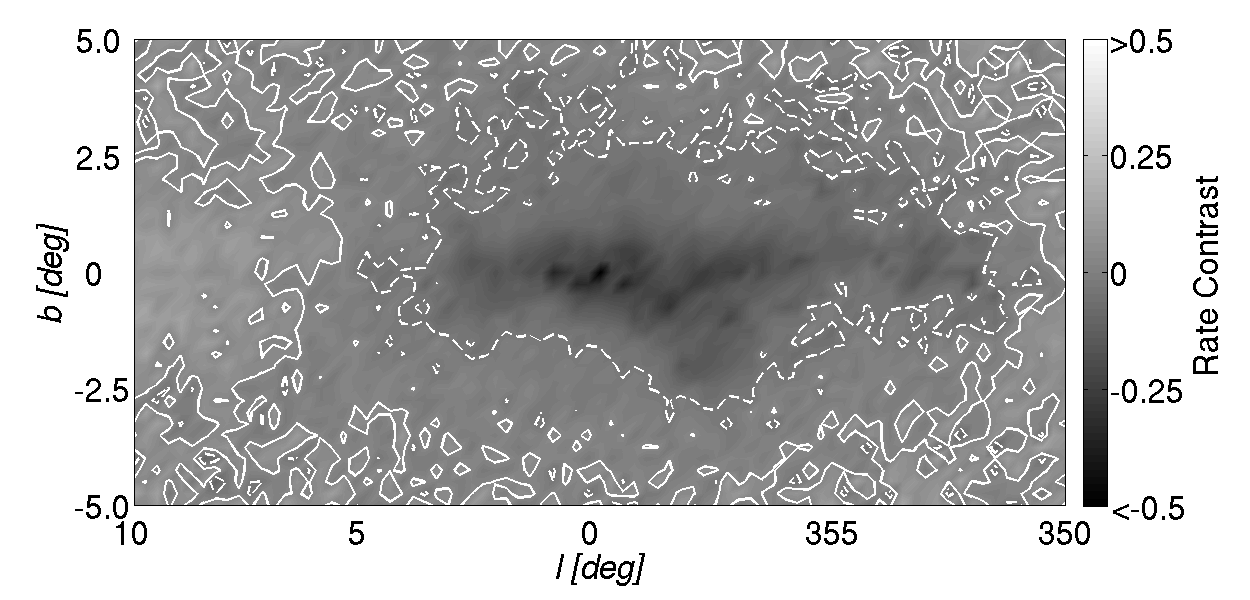} & \hspace{-0.3in}
\includegraphics[width=3.4in]{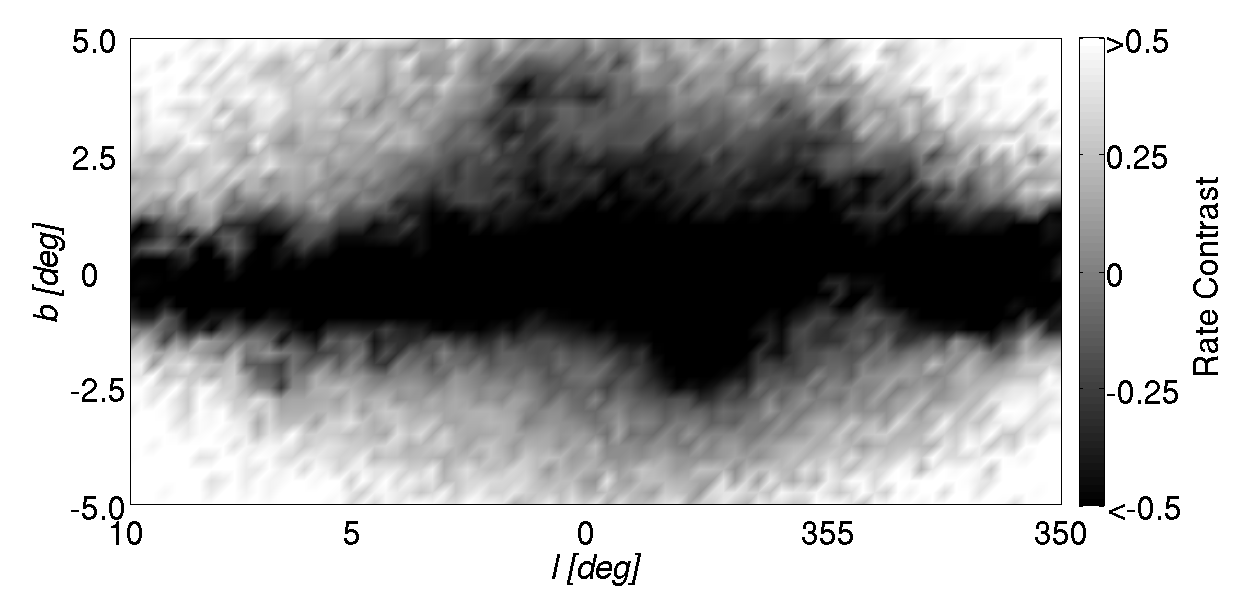} \\
Space-based $H$-band, Jupiter-sized FFP & Space-based $H$-band, Neptune-sized FFP & Space-based $H$-band, Earth-sized FFP\\
\includegraphics[width=3.4in]{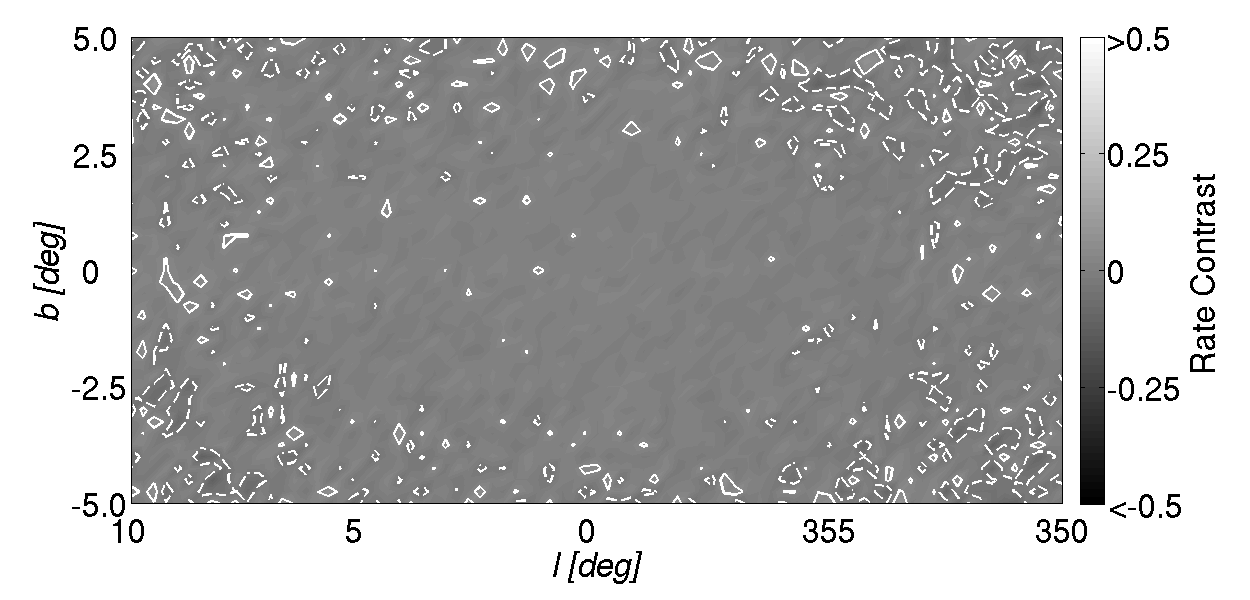} & \hspace{-0.3in}
\includegraphics[width=3.4in]{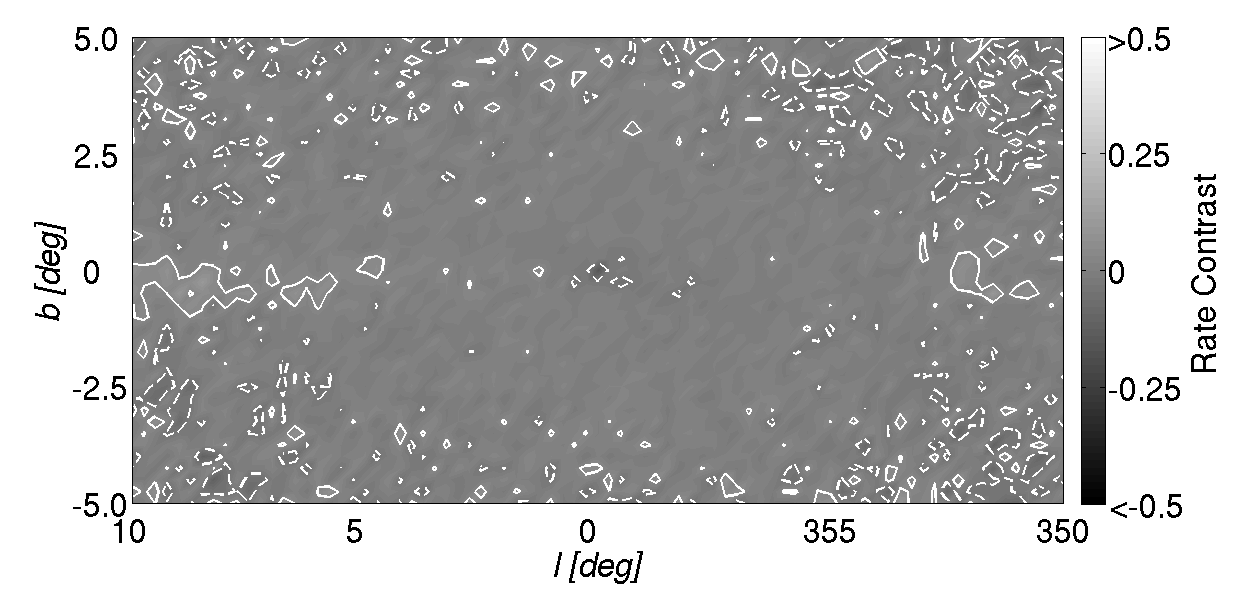} & \hspace{-0.3in}
\includegraphics[width=3.4in]{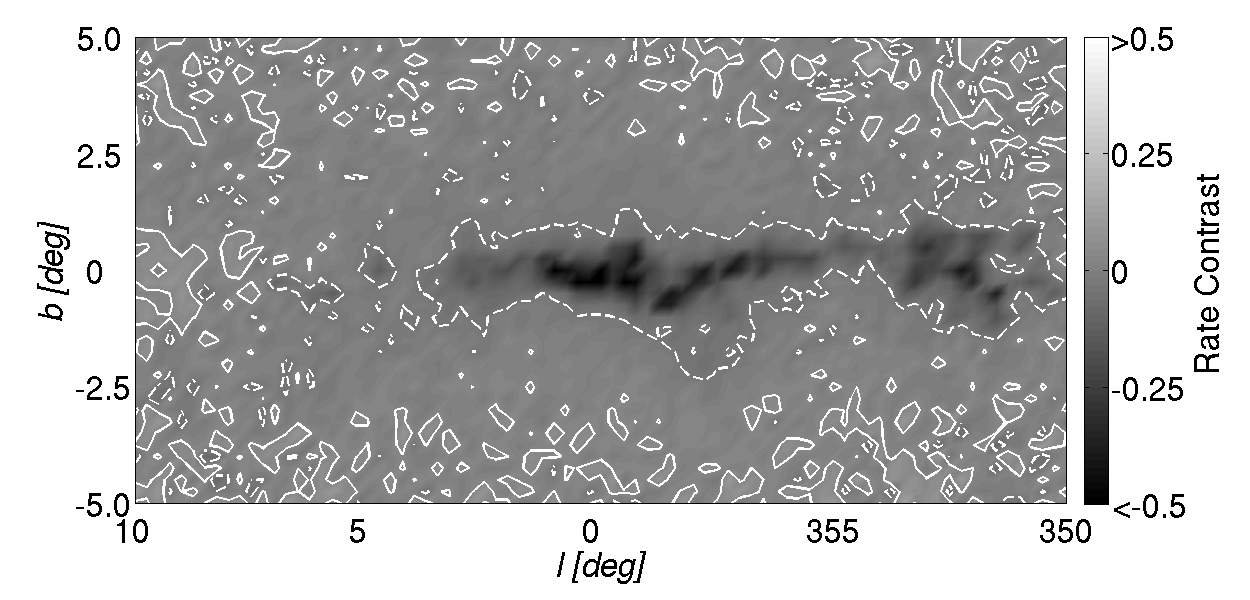} \\
\end{tabular}
}
\caption{Maps of the event rate contrast, RC (see main text). The columns are Jupiter-mass, Neptune-mass, and Earth-mass FFPs from left to right. The rows are ground-based $I$-band, ground-based $K$-band, and space-based $H$-band observations from top to bottom. For the grey monotone maps, the solid (RC $= 0.03$) and dashed ($-0.03$) contours are drawn. In maps exhibiting very large RC variation these contour levels are seen to be nearly coincident.}
\label{fig:rc-ffp-map}
\end{figure}
\end{landscape}

The sensitivity is brought out more clearly by dividing the FFP rate maps by the stellar event rate maps shown in Figure \ref{fig:er-star-map}. Maps of the deviation of this ratio from the global median ratio are shown in Figure \ref{fig:rc-ffp-map}. We refer to the deviation as the event rate contrast (RC), where
\begin{equation}
  \mbox{RC} = (\Gamma_{\rm FFP}/\Gamma_*) - \mbox{median}(\Gamma_{\rm FFP}/\Gamma_*)
\end{equation}
and subscripts denote FFP and stellar ($*$) event rates. Low variation in the RC map implies that the FFP rate traces the ordinary stellar microlensing rate. This is a desirable property if, for example, a restricted area is being used to try to determine the global abundance of FFPs. Large variation in the RC map means that the target area would require careful selection to remove spatial bias in the derived occurrence rate. The spatial distribution of FFPs may be sensitive not only to the FFP mass scale, but also to the details which go into the Galactic model, especially the 3D modelling of the dust distribution. Figure \ref{fig:rc-ffp-map} shows that there is significant variation in RC for ground-based $I$-band surveys towards the Galactic plane. In practice it may be hard to infer global FFP statistics from ground-based observations without statistically meaningful samples of FFPs away from regions of significant extinction. This is true even for $K$-band surveys where dust is less problematic for stellar microlensing. The results for the space-based simulation show much less sensitivity to the FFP mass scale and essentially show that the FFP rate traces the stellar rate across virtually the whole of the inner Galaxy. This means that global statistics about the FFP distribution can be reliably determined for space-based surveys which may only target a relatively small area and that the result should not depend critically on the location of the survey.

\begin{table}
 \centering
  \caption{The FFP event rate $\Gamma\times N_*$ [events yr$^{-1}$ deg$^{-2}$] at $(l, b) = (1^{\circ}, -1.^{\circ}75)$ in SN mode for the limit of $100\%$ survey detection efficiency. The rate contrast is given in square brackets. A correction factor of 1.6 has been applied.}
  \label{tab:penny-ffp}
  \begin{tabular}{lllll}
  \hline
  Lens mass & $I$ band & $K$ band & $H$ band\\
  & (Ground) & (Ground) & (Space)\\ \hline \hline
  Jupiter & 180 [0.8\%] & 470 [-0.2\%] & 620 [0.6\%]\\
  Neptune & 29 [-2.2\%] & 98 [-1.4\%] & 140 [0.5\%]\\
  Earth   & 0.6 [-53\%] & 10 [-18\%]   & 33 [0.8\%]\\	\hline
\end{tabular}
\end{table}

The maximum event rate and the area integrated total rate for each map are summarised in Table \ref{tab:maxer-ffp}; the factor 1.6 scaling is applied because the model underpredicts the rate (see Section \ref{subsubsec:od-star}). The table statistics are summarised for a hypothetical benchmark of continuous monitoring throughout the year without interruption over the entire simulated survey area. We note that the numbers reflect the expected yield for a survey with $100\%$ detection efficiency across all timescales. It should also be noted that proposed space-based microlensing surveys intend only to cover a few square degrees of sky and so the ``max'' rate is more indicative of space-based performance than the ``global'' rate.

Table \ref{tab:penny-ffp} shows stellar and FFP rate predictions at ($l, b$) = ($1^{\circ}, -1.^{\circ}75$), approximately at the central target position of the proposed ExELS exoplanet microlensing survey using \euclid{} \cite{penny13}. \cite{penny13} performed a detailed detection simulation for the ExELS exoplanet additional science survey proposal for \euclid{}. They find that the stellar microlensing rate is $\sim 27,000$ events/yr within the ExELS fields for events with $u < u_{\max} = 3$. By comparison, we find an event rate of $\sim 43,000$ events/yr over the same region based on the rate at the central location of the ExELS fields. \cite{penny13} predict four Earth-mass FFPs with $u < u_{\max} = 1$ would be detected by \euclid{} for a survey area of 1.6 deg$^2$ over a 300-day observation baseline. They assume one FFP per Galactic star, which is the same assumption as used for our predictions. Our space-based $H$-band rate corresponds to $\sim$43 Earth-mass FFPs with $u< u_{\max} = 1$ during the \euclid{} mission . This is one order of magnitude larger than the Penny et al. result, though as emphasised previously, we assume 100\% detection efficiency across all timescales. \cite{penny13} take some account of detection efficiency through cuts on the minimum number of data points through the FFP event as well as a cut on the microlensing goodness of fit to their simulated photometry. This is a significantly more stringent requirement than our cut on the peak theoretical signal-to-noise ratio. Actual survey yields are therefore expected to be an order of magnitude below our theoretical yields.

\subsubsection{Space-based versus ground-based rates} 

Whilst our general simulations do not take account of survey-specific efficiencies we can partially account for some inefficiencies in the observing strategies of space- and ground-based observations.
 
For ground-based surveys, the bulge is not observable throughout the year and observations are also subject to weather interruptions. \cite{hend14} compute the effective bulge observability at the three nodes of the KMTNet survey. Their Figure 2 indicates that the bulge is observable for at least one of these telescopes for, on average, around 30\% of the time over 8 months of the year. This implies an observing efficiency of approximately $20\%$ averaged over a whole year. For single-site surveys such as MOA and OGLE we can expect an efficiency nearer to $10\%$, depending on weather. 
Neither current ground-based nor proposed space-based surveys cover our full simulated area with an observing cadence necessary for FFP detection.
The OGLE-IV survey monitors nine fields with a cadence of higher than three observations per night over a collective area of around 11 deg$^2$. MOA-2 monitors around 13 deg$^2$ with better than one-hour cadence. For KMTNet \cite{hend14} advocate a high-cadence coverage of around 16 deg$^2$ as being optimal for the discovery of low-mass planets.

Proposed space-based surveys with Euclid and WFIRST are not restricted by weather but cannot compete with the areal coverage achievable from the ground. They are also restricted by their orientation with respect to the Sun. In practice, this restricts the total time for which they can conduct bulge microlensing programmes as the bulge lies close to the ecliptic plane. For this reason the maximum lifetime of microlensing programmes is typically of the order of one year of on-bulge time, but spread over several years of the mission. Table \ref{tab:survareas} lists the areal coverage and survey efficiency for a number of current and proposed ground and space microlensing surveys. For the space missions (Euclid and WFIRST) the efficiency in Table \ref{tab:survareas} is simply given by the likely total on-bulge observing time of the proposed programme. For the ground-based surveys the efficiency represents the fraction of the yearly rate likely to be achievable within a given observing season.

\begin{figure}
\centering{
\includegraphics[width=4in]{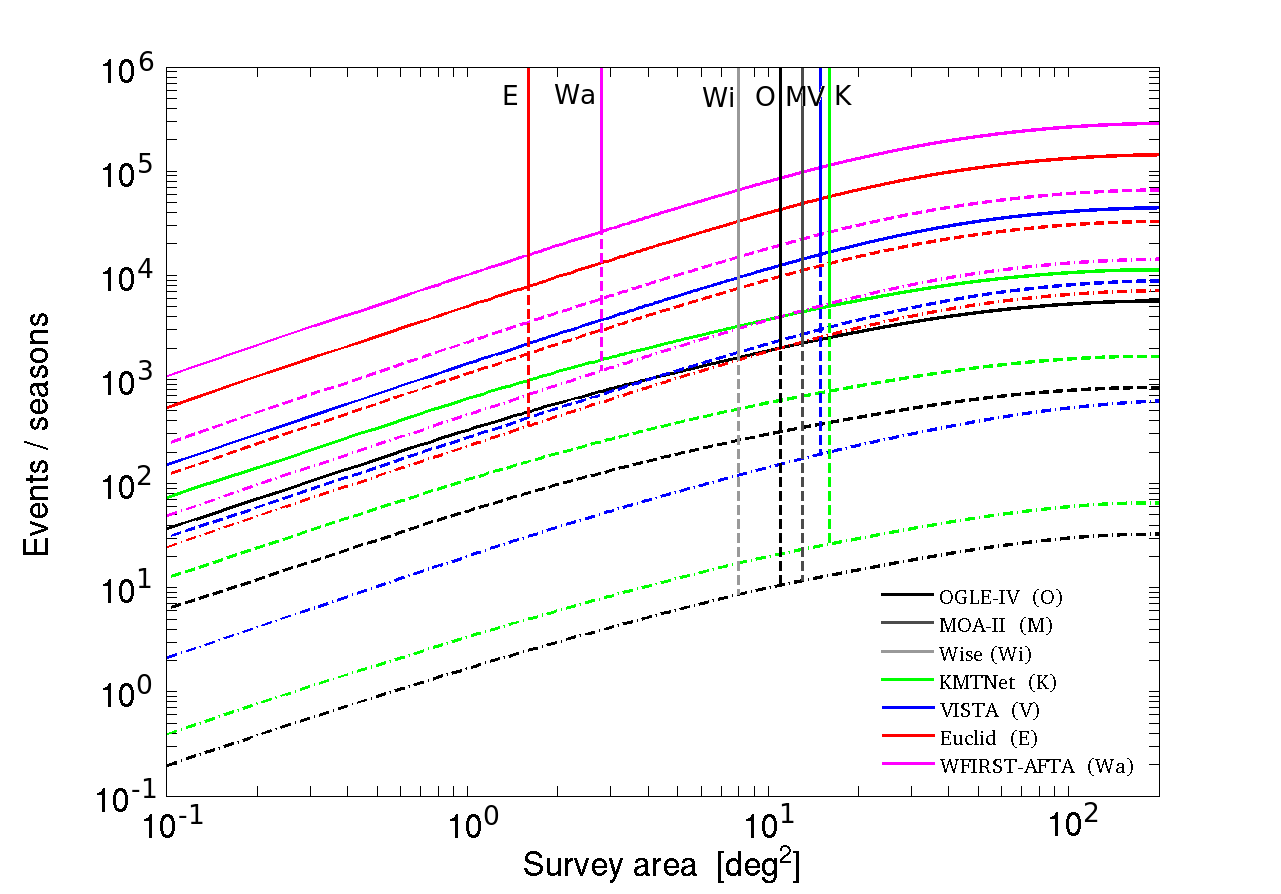}
}
\caption{Optimised event rate versus survey area for FFPs of Jupiter mass (solid lines), Neptune mass (dashed lines) and Earth mass (dot-dashed lines). The survey area is assumed to be distributed around the regions of maximum yield for that area, and therefore do not necessarily involve contiguous survey areas. Yields are indicated for several current and proposed ground- and space-based surveys, assuming the areas and efficiencies listed in Table \ref{tab:survareas}. For the space-based programmes the number of events refers to the yield for the whole proposed observation baseline, whilst for other surveys the number of events refers to a single observing season.}
\label{fig:survyield}
\end{figure}

The numbers in Table \ref{tab:survareas} can be used to scale the rate yields in Table \ref{tab:maxer-ffp} to approximate expected seasonal rates for a given survey in the limit of perfect detection capability across all event timescales. Figure \ref{fig:survyield} shows the maximum event rate detectable as a function of survey area for ground and space surveys listed in Table \ref{tab:survareas}. The actual area of the high-cadence regions of each survey is indicated on each line. We note that these maximum rates are obtained by rank sorting regions of the rate maps by the highest yielding regions and therefore a given total areal coverage may correspond to non-contiguous regions which do not correspond to the actual area being surveyed by the surveys. They instead represent the largest event yield which can be obtained for a survey with the same total areal coverage, wavelength and sensitivity. 

The advantage of a multi-site ground-based survey like KMTNet over single-site surveys is apparent from Figure \ref{fig:survyield}. However a high-cadence near-IR survey with VISTA covering a similar area could prove even more effective owing to the increased sensitivity of near-IR observations. Figure \ref{fig:survyield} shows how space-based surveys are particularly effective for detecting low-mass exoplanets. Even though Euclid and WFIRST will cover areas of only around a couple of square degrees, their detection yield for Earth-mass FFPs is expected to be 1--2 orders of magnitude higher than ground-based survey seasonal rates which span an area of sky 10 times larger. For Jupiter-mass planets the space mission yields are 10--30 times higher, and so are still likely to exceed yields from the ground, with the possible exception of a high-cadence VISTA survey operating over at least a decade. Overall, a short space-based campaign over a modest area can significantly outperform a gtound-based campaign and can provide important statistics on the FFP population over a range of FFP mass.

In our simulation, the population of one FFP per star is assumed, but as \cite{sumi11} mentioned, the population of planets may be larger than stars. The detection probability of FFP is therefore expected to rise from our event rate values. Moreover, \cite{varvoglis12} and \cite{wang15} calculated the motion of FFPs in a cluster and resulted they tend to travel within the cluster unless any outside effect such as a close encounter with other clusters happens. The capture and release of FFPs (\citealt{varvoglis12}) may also temporarily and slightly affect the population. Thus, the FFP distribution is similar to the stellar distribution, but may be larger and sometimes fluctuate.

\begin{table*}
 \centering
 \begin{minipage}{140mm}
  \caption{Areal coverage and efficiency of current and proposed surveys capable of detecting FFPs. The areal coverage is determined by the area in which high-cadence observations are being undertaken (\citealt{udalski15}, \citealt{sumi10}, \citealt{mainzer05}, \citealt{hend14}, \citealt{hempel14}, \citealt{penny13}, \citealt{spergel15}). For ground-based, the efficiency gives the fraction of the yearly rate which is detectable due to the year-round observability of the bulge and weather interruptions. For space-based, the efficiency indicates the total baseline of proposed exoplanet survey, expressed as a fraction of a year.} 
\label{tab:survareas}
  \begin{tabular}{lllll}
  \hline
   Survey & Location & Filter & High-cadence & Efficiency \\
          &          &        & area (deg$^2$) & \\
   \hline
   OGLE-IV & Chile & $I$ & 11 & 0.1 \\
   MOA-II & Australia & $I$ & 13 & 0.1 \\
   Wise & Israel & $I$ & 8 & 0.1 \\
   KMTNet & Chile/South Africa/Australia & $I$ & 16 & 0.2 \\
   VISTA & Chile & $K$ & 15 & 0.1 \\
   Euclid & Space & $H$ & 1.6 & 0.2 \\
   WFIRST-AFTA & Space & $H$ & 2.8 & 0.4 \\
   \hline
\end{tabular}
\end{minipage}
\end{table*}

\section{Discussion and conclusion}	\label{sec:conc}		

We have performed a detailed simulation of microlensing by free-floating planets (FFPs) using an updated version of the \bes{} population synthesis model of the Galaxy which incorporates a two-component bulge \cite{robin12}. We have simulated signal-to-noise limited maps of the optical depth, event rate and average event duration for stars and for FFPs of Jupiter, Neptune and Earth masses. Motivated by the tentative observation of \cite{sumi11} we  assume one FFP per Galactic star. Our simulations cover an optical ground-based survey with characteristics similar to MOA-II and OGLE-IV, a deep near-infrared $K$-band survey, and a space-based $H$-band survey motivated by the ExELS proposal for the \euclid{} mission \cite{penny13}. Our simulation includes a 3D dust model and takes into account of finite source size effects for FFP microlensing.

Using the predictions for stellar lensing we determine a correction factor of 1.6 to scale the \bes{} model rate up to the event rate observed by the MOA-II survey \cite{sumi16}, similar to the approach of other studies \cite[e.g.][]{penny13,awip16}. 

For FFPs we find that the average duration towards the Galactic centre is modified by finite source size effects with lower mass FFPs lasting a little longer than under point source assumptions.
The potential ground-based $I$-band FFP theoretical event rate across the whole bulge region is as large as 3,100 events/yr if there is a Jupiter-mass FFP for every Galactic star, though the event rate observed in practice from a real survey is expected to be an order of magnitude below the theoretical rate. For Earth-mass FFPs with the same abundance the theoretical rate is 18 events/yr. These numbers assume $100\%$ detection efficiency for a survey monitoring 200 deg$^2$ with temporal sensitivity capable of detecting very short duration events, and are therefore much larger than the KMTNet simulation performed by \cite{hend14}. They simulate a baseline survey over 16 deg$^2$ (rather than 200 deg$^2$) and include selection cuts on simulated photometry, which inevitably reduces the actual yield with respect to the theoretically achievable yield computed here for an idealised survey.

We find that our simulated VISTA-like survey provides an FFP yield which is a factor $> 4$ times higher than optical surveys for the same FFP mass owing to the lessened extinction effects. A space-based $H$-band survey has two orders of magnitude greater sensitivity to Earth-mass FFP detection than optical ground-based surveys. However, such a survey will necessarily be conducted over a restricted area of sky so that the actual gain for an areal coverage of a few deg$^2$ will be closer to one order of magnitude for Earth-mass FFPs. The expected yields for higher mass FFPs  will be similar to large-scale ground-based survey yields. The ground-based survey sensitivity falls off substantially at Earth-mass scales owing to finite source size effects.

FFP microlensing surveys aim to understand the global population of FFPs and therefore need to ensure that any spatial biases are understood. We find that the combination of Galactic dust and finite source size effects play an important role here. Both $I$- and $K$-band ground-based survey simulations show that the spatial distribution of FFPs does not follow that of the stars within the inner Galaxy, but is sensitive to the dust distribution. The spatial distribution of FFPs also varies with FFP mass. At Earth-mass scales the FFP spatial distribution is very sensitive to the dust map and therefore to the assumed Galactic model and this may make it difficult to measure accurately the FFP population abundance. Our space-based $H$-band simulations show a much more uniform FFP distribution which traces the stellar microlensing rate over almost the entire simulation area and over all simulated FFP mass scales. This means that spatial bias should not be a critical factor in the choice of location for space-based FFP surveys. It may also mean that space-based surveys offer the most reliable determination of FFP abundance in that they depend minimally on fine details of the 3D dust distribution.

\bibliographystyle{aa}
\bibliography{ref}

\end{document}